%% file: elsarticle-template-num.tex
\documentclass[final,5p,times,twocolumn]{elsarticle}

\usepackage{amssymb}
\usepackage[utf8]{inputenc} % allow utf-8 input
\usepackage[T1]{fontenc}    % use 8-bit T1 fonts
\usepackage{hyperref}       % hyperlinks
\usepackage{url}            % simple URL typesetting
\usepackage{booktabs}       % professional-quality tables
\usepackage{tikz}
\usepackage{pgfplots}
\pgfplotsset{compat=1.16}
\usepackage{amsmath,amsfonts, amsthm}
\usepackage{bm}
\usepackage{accents}
\usepackage{nicefrac}       % compact symbols for 1/2, etc.
\usepackage{microtype}      % microtypography
\usepackage{lipsum}
\usepackage{caption}
\usepackage{subcaption}
\usepackage[short, nocomma]{optidef}
\usepackage{fancyhdr}       % header
\usepackage{graphicx}       % graphics
\usepackage{float}
\graphicspath{{media/}}     % organize your images and other figures under media/ folder
\usepackage{algorithm}
\usepackage{algpseudocode}
\newcommand{\ubar}[1]{\underaccent{\bar}{#1}}

\newcommand{\red}[1]{{\color{black}#1}}

\newtheorem{remark}{Remark}
\newtheorem{lemma}{Lemma}
\newtheorem{assumption}{Assumption}
\newtheorem{definition}{Definition}

\usepackage{framed} % Framing content

\usepackage{multicol} % Multiple columns environment

\usepackage{nomencl} % Nomenclature package

\makenomenclature

\renewcommand*\nompreamble{\begin{multicols}{2}}

\renewcommand*\nompostamble{\end{multicols}}
\journal{Sustainable Energy, Grids and Networks}

\begin{document}

\begin{frontmatter}



\title{Risk-aware Scheduling and Dispatch of Flexibility Events in Buildings}


\author[inst1,inst2]{Paul Scharnhorst\corref{cor1}}\ead{paul.scharnhorst@csem.ch}
\cortext[cor1]{Corresponding author}
\affiliation[inst1]{organization={Digital Energy Solutions Group, CSEM SA},%Department and Organization
            %addressline={Rue Jaquet Droz 1}, 
            city={Neuchâtel},
            postcode={2000}, 
            %state={State One},
            country={Switzerland}}
\affiliation[inst2]{organization={Automatic Control Laboratory, EPFL},%Department and Organization
            %addressline={Address Two}, 
            city={Lausanne},
            postcode={1015}, 
            %state={State Two},
            country={Switzerland}}
\author[inst1]{Baptiste Schubnel}\ead{baptiste.schubnel@csem.ch}
\author[inst1]{Rafael E. Carrillo}\ead{rafael.carrillo@csem.ch}
\author[inst1]{Pierre-Jean Alet}\ead{pierre-jean.alet@csem.ch}
\author[inst2]{Colin N. Jones}\ead{colin.jones@epfl.ch}

\begin{abstract}
Residential and commercial buildings, equipped with systems such as heat pumps \red{(HPs)}, hot water tanks, or stationary energy storage, have a large potential to offer their consumption flexibility as grid services. In this work, we leverage this flexibility to react to consumption requests related to maximizing self-consumption and reducing peak loads. We employ a data-driven \red{virtual} \red{storage} modeling approach for flexibility prediction in the form of flexibility envelopes for individual buildings. The risk-awareness of this prediction is inherited by the proposed scheduling algorithm. A Mixed-integer Linear Program (MILP) is formulated to schedule the activation of a pool of buildings in order to best respond to an external aggregated consumption request. This aggregated request is then dispatched to the active individual buildings, based on the previously determined schedule. The effectiveness of the approach is demonstrated by coordinating up to 500 simulated buildings using the Energym Python library and observing about 1.5 times peak power reduction in comparison with a baseline approach while maintaining comfort more robustly. We demonstrate the scalability of the approach by solving problems with 2000 buildings in about 21 seconds, with solving times being approximately linear in the number of considered assets.
\end{abstract}




\end{frontmatter}


\section{Introduction}
Demand Response (DR) has been identified as a key technology to help balance the power grid under growing shares of renewable energy sources and increased electrification \cite{WEO_IEA}. Among the flexible assets able to provide DR services, both residential \cite{resflexrev} and commercial \cite{comflexrev} buildings have been recognized as promising candidates to provide services such as peak reduction through flexible use of, e.g., their Heating, Ventilation, and Air Conditioning (HVAC) systems.

% At this stage it would be helpful to define the components or the actors of a demand response scheme, as it would make it easier to understand the needs and limitations that are discussed in the bulk of the introduction. This could be done e.g., by referring to USEF and by bringing upwards fig:approach, whose content is relatively general.

The viewpoint taken in this work is in line with the Universal Smart Energy Framework (USEF) \cite{2021usef}: Small to medium-sized assets, referred to as Active Customers in the USEF, provide the consumption flexibility that is coordinated by an aggregator. This aggregator interacts with the market to sell the flexibility to different parties while guaranteeing the delivery of the requested flexibility despite uncertainties. The approach proposed in this work is especially aimed at aggregators interacting with individual assets.

To effectively use the consumption flexibility of different assets, two requirements are needed: The quantification of available flexibility and the coordination of the different assets to ensure optimal dispatch.

For the first requirement, approaches either aim at quantifying the aggregate flexibility of a pool of assets or the flexibility of individual assets. An overview of different quantification methodologies is given in \cite{REYNDERS2018review}. Aggregate estimation is discussed, e.g., in the works \cite{FISCHER2017853, YIN2016149, JUNKER2018175}. Many approaches deal with the estimation of individual flexibility, either in a model-based \cite{gasser_predictive_2021} or data-driven \cite{hekmat2021datadriven} way. Popular is also the approach of virtual-battery modeling for thermal assets, as presented in e.g. \cite{hao2013, hao2015, sanandaji2014} for Thermostatically Controlled Loads (TCLs). Since considering uncertainty is vital to guaranteeing comfort when dealing with systems that provide heating or cooling \cite{OLDEWURTEL201215}, we apply the approach presented in \cite{batmodel} in this work to quantify available flexibility. \red{Recent studies investigated similar methods to quantify flexibility while being uncertainty-aware \cite{10407489}.} 

However, the focus of this work lies on the second requirement, the coordination of flexibility in a pool of assets. This problem has been addressed in a number of works. Hierarchical control approaches have been proposed in, e.g., \cite{BORSCHE201410299}, \cite{vrettos2016}, and \cite{QURESHI2018216}, where the former uses a scenario-based approach and the latter two employ robust control methods. The principle of these hierarchical controllers is to have a high-level mechanism to plan available reserves in advance, e.g. in a day-ahead fashion, an intermediate level MPC scheme to optimize operation while keeping the planned reserves, and a low-level controller to track the reference determined by the MPC. However, these solutions require accessibility to low-level control, which can be prohibitive for some users. A priority-stack-based controller for switching TCLs, similar to the ones used in the previously mentioned works on \red{virtual} battery modeling for TCLs, is used in \cite{mathieu2013}, considering different levels of available information in the control. 

Scheduling-based approaches for the operation of flexible resources have been presented in multiple works. \cite{10.1007/978-3-319-23868-5_16} considers time shiftable and power shiftable devices in a scheduling problem for a whole district. With the aim of flattening the consumption profile of that district, the authors use a meta-heuristic approach to deal with the high number of variables in the scheduling problem. Considering different loads in a single household, \cite{6509468} proposes a MILP to schedule appliance operation for cost minimization with respect to an external signal. While targeting individual buildings equipped with a device for heuristically solving the scheduling, called Energy Box, the authors argue that the approach is also applicable to a group of buildings. Flexibility of appliances in residential and commercial buildings is also leveraged in the scheduling approach presented in \cite{AYON20171}. It relies on aggregating the available flexibility before the scheduling and disaggregating it later. However, a simulation tool is needed to quantify the available flexibility by simulating the assets with varying setpoints. Minimizing operation cost is also the objective in \cite{OTTESEN2015364} where the authors consider systems with different constraints in their scheduling problem, thus modeling, e.g., shiftable profiles or curtailable profiles. To handle uncertainty, rolling horizon versions of the approach are proposed, either in a deterministic fashion or in a stochastic fashion. \red{Recent studies address similar settings, with \cite{smartresidential} using MILP to minimize grid imports for appliances, battery storage, and EVs, or \cite{anoveladaptive} proposing a bi-level optimization problem that handles uncertainties of consumption and PV production for cost minimization.} The different works presented in this paragraph all consider direct cost minimization as their objective and therefore do not consider request following as an application. \cite{OLIVELLAROSELL2018881} extends the approach from \cite{OTTESEN2015364} to include consumption requests in the constraints. Most of the mentioned works rely on some form of explicit modeling of systems or their flexibility.

Multiple optimization-based approaches to determine the activation of flexible assets and generation units are proposed for the unit commitment problem. An example of this is given in \cite{bertsimas2013uc}, where a robust approach to deal with uncertainty in large-scale mixed-integer programs is presented. Different types of uncertainty are considered in \cite{Velloso_2020}, employing a dynamic uncertainty set that is adapted based on observations. A risk measure approach using Conditional Value at Risk (CVaR) is used to avoid conservativeness in robust approaches due to rare events in \cite{Kazemzadeh2019}. Other methods that consider DR schemes as inputs to their scheduling problems are given in \cite{Ghahramani2019, su142114194}.

Based on these considerations, we propose a method to coordinate a pool of assets to react to external consumption requests, focusing on the cases of self-consumption and peak reduction. Inheriting the uncertainty-aware estimation from \cite{batmodel}, our scheduling approach works with an intuitive description of flexibility that can cover many different assets without changing the scheduling problem formulation. Our contributions are as follows:
\begin{itemize}
    \item We propose an efficient way to compute uncertainty-aware flexibility envelopes based on the virtual \red{storage} model presented in \cite{batmodel} (therein called "virtual battery model")
    \item We present a scheduling problem formulation that is compliant with general request following and demonstrate the objectives of self-consumption maximization and peak reduction
    \item We demonstrate the effectiveness and scalability of the approach in a large-scale simulation study
\end{itemize}

\begin{figure*}
    \centering
    
    \includegraphics[width=\textwidth]{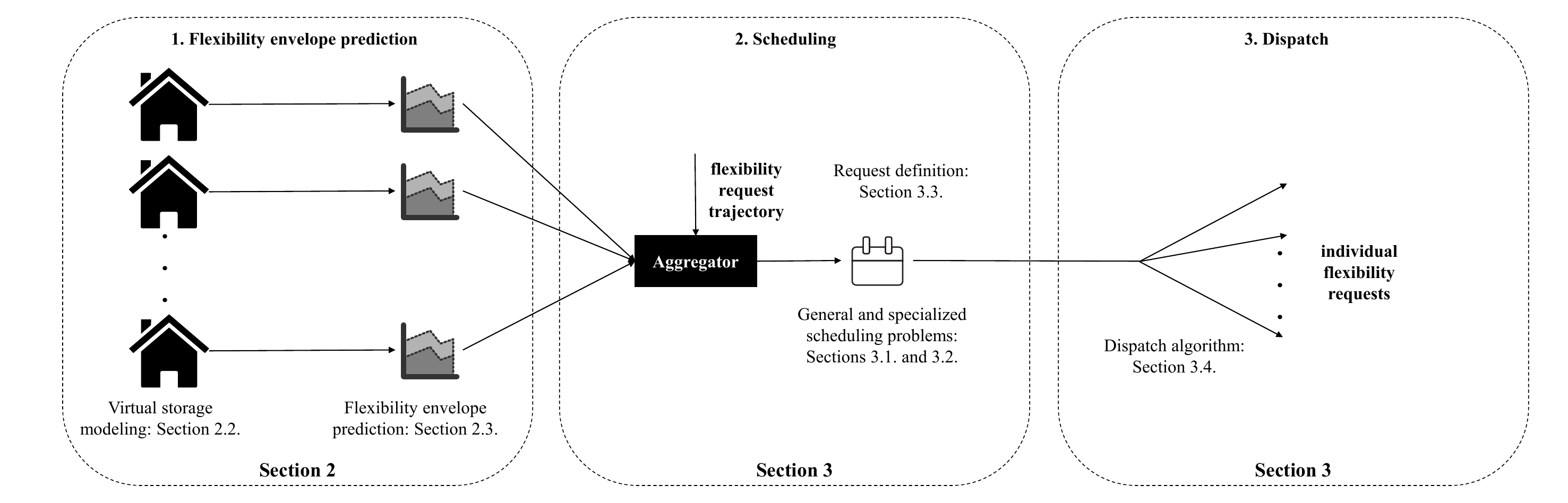}
    %\makebox[\textwidth][l]{\input{scheme}}
    %\vphantom{\includegraphics[width=0.7\textwidth]{figs/Slide1.png}}
    %\includegraphics[width=0.7\textwidth]{figs/Slide1.png}%{figs/scheme_dispatch_paper.pdf}
    \caption{The three phases of our proposed approach.}
    \label{fig:approach}
\end{figure*}

%First, we provide a flexibility definition for individual assets and a systematic way to estimate it in a data-driven and uncertainty-aware way, using the results from \cite{batmodel}. Second, we formulate scheduling problems using MILPs to determine the activation times of a pool of assets to best follow a consumption request trajectory. The exact dispatch of incoming requests between the active assets of the pool is done via another optimization or a heuristic algorithm. Third, we demonstrate the effectiveness of our approach in a large-scale simulation study, comparing the performance to a baseline approach and investigating the influence of the chosen risk level. 
A \red{flowchart} of the flexibility prediction, scheduling, and dispatch of our approach with corresponding sections is shown in Figure~\ref{fig:approach}.

%\begin{figure}
    %\centering
    
    %\input{figures/scheme}
    %%\makebox[\textwidth][l]{\input{scheme}}
    %%\vphantom{\includegraphics[width=0.7\textwidth]{figs/Slide1.png}}
    %%\includegraphics[width=0.7\textwidth]{figs/Slide1.png}%{figs/scheme_dispatch_paper.pdf}
    %\caption{The three phases of our proposed approach.}
    %\label{fig:approach}
%\end{figure}

The paper is structured as follows. Section~\ref{sec:flex_est} introduces the flexibility characterization used in this work, in the form of flexibility envelopes. It also recaps a data-driven virtual \red{storage} modeling approach for buildings from \cite{batmodel}, used for predicting uncertainty-aware flexibility envelopes, which are employed in scheduling individual buildings for activation. This scheduling problem is presented in Section~\ref{sec:sched}, with specific formulations for the cases of self-consumption and peak power reduction. Additionally, an algorithm for dispatching aggregated consumption requests, based on the determined schedules, is introduced. Section~\ref{sec:exp} explains the details of the conducted simulation study, including the type of the consumption requests, the baseline approach that serves as a comparison, and the considered evaluation metrics. The results are then presented and discussed in Section~\ref{sec:res} before the paper is concluded in Section~\ref{sec:con}.

\begin{table*}[!t]   
\red{
\begin{framed}

\nomenclature{$\bm{x}_t$}{General state variable}
\nomenclature{$h$}{State update function}
\nomenclature{$p_t$}{Power input}
\nomenclature{$\bm{e}_t$}{External conditions}
\nomenclature{$\omega_t$}{Unmeasured disturbances}
\nomenclature{$\ubar{\bm{x}}, \bar{\bm{x}}$}{General state bounds}
\nomenclature{$\ubar{p}, \bar{p}$}{Power limits}
\nomenclature{$\bm{p}^b$}{Baseline power trajectory}
\nomenclature{$r_t$}{Relative consumption request}
\nomenclature{$\bm{R}^k_{0:H-1}$}{Fixed-time flexibility envelope}
\nomenclature{$s_t$}{Lumped state variable}
\nomenclature{$\hat s_t$}{Approximation of lumped state variable}
\nomenclature{$a^+, a^-$}{Model parameters (random variables)}
\nomenclature{$\mathcal{P}^+, \mathcal{P}^-$}{Sample spaces of $a^+, a^-$}
\nomenclature{$b_f$}{Model parameter}
\nomenclature{$f$}{Baseline state model}
\nomenclature{$\chi_r$}{Indicator function}
\nomenclature{$\alpha$}{Uncertainty parameter}
\nomenclature{$\bm{r}^{\text{agg}}$}{Aggregate request trajectory}
\nomenclature{$u_{i,t}$}{Activation variable of asset $i$}
\nomenclature{$\epsilon$}{Request covering parameter}
\nomenclature{$d_t$}{Pointwise scaling factor}
\nomenclature{$\bm{r}^{\text{comm}}$}{Committed request trajectory}
\nomenclature{$\bm{g}^{\text{agg}}$}{Aggregate production trajectory}
\nomenclature{$\bm{r}^{\text{self}}$}{Self-consumption request trajectory}
\nomenclature{$c$}{Requested peak consumption}
\nomenclature{$\bm{r}^{\text{peak}}$}{Peak reduction request trajectory}
\nomenclature{$\rho$}{Achievable peak consumption variable}
\nomenclature{$A_{i,t}$}{Activation variable of building $i$}
\nomenclature{$F_{i}$}{Flexibility potential of building $i$}
\nomenclature{$\Delta P_a$}{Absolute peak power reduction}
\nomenclature{$g_{\text{sum}}$}{Total summed production}
\nomenclature{$\Delta S_r$}{Self-consumed power fraction}
\nomenclature{$\ubar{T}, \bar{T}$}{Temperature bounds}
\nomenclature{$\Delta T_r$}{Percentage of temperature bound violations}

\printnomenclature

\end{framed}}

\end{table*}

\section{Flexibility estimation}
\label{sec:flex_est}
This section introduces the consumption flexibility characterization later used in the scheduling of the assets. A short overview of a method to predict uncertainty-aware flexibility envelopes, using a risk-aware, data-driven virtual \red{storage} model, is given. Furthermore, we extend the results of \cite{batmodel} here by proposing an efficient and scalable approach to compute the flexibility envelopes.

The flexibility envelopes are generic and can be applied to different assets. In the numerical experiments of this paper, we focus on building assets with HPs to provide consumption flexibility. 

\subsection{Characterization of flexibility}
We consider an asset with state dynamics given by
\begin{equation}
\label{eq:dyngen}
    \bm{x}_{t+1} = h(\bm{x}_t, p_t, \bm{e}_t, \omega_t)
\end{equation}
where $\bm{x}_t \in \mathbb{R}^{n_x}$ denotes a collection of states and measured variables at time $t$, for example, the temperature in thermal assets, $p_t\in \mathbb{R}$ the power input, $\bm{e}_t \in \mathbb{R}^{n_e}$ a collection of external conditions, for example, outdoor temperature or solar irradiance, and $\omega_t \in \mathbb{R}^{n_{\omega}}$ unmeasured disturbances, like internal gains through occupants in buildings. $h:\mathbb{R}^{n_x}\times \mathbb{R} \times \mathbb{R}^{n_e} \times \mathbb{R}^{n_{\omega}}\to \mathbb{R}^{n_x}$ denotes the function describing the state transition.

These assets are subject to constraints, where in this work we consider box constraints of the form
\begin{equation}
    \label{eq:consbdstategen}
    \ubar{\bm{x}} \leq \bm{x}_t \leq \bar{\bm{x}}
\end{equation}
with the lower and upper bounds denoted by $\ubar{\bm{x}}, \bar{\bm{x}}\in \mathbb{R}^{n_{x}}$ and the inequalities applied entry-wise. These bounds can be temperature bounds in thermal assets or bounds on the state of charge in batteries. Additionally, the asset might be subject to power constraints \begin{equation}
\label{eq:conspw}
    \ubar{p}\leq p_t \leq \bar{p}.
\end{equation} Ramp-rate constraints are not considered in this work but can be incorporated in a characterization of flexibility very similar to the one presented here.

As a last preliminary step, we introduce the notion of \emph{relative consumption requests} with respect to a baseline consumption trajectory. This baseline is due to the following assumption:
\begin{assumption}
    The considered asset is actively controlled. The baseline trajectory of power inputs resulting from the nominal controller operation is denoted by $\bm{p}_{0:H-1}^{\text{b}} = [p_{0}^{\text{b}}, \dots, p_{H-1}^{\text{b}}]^\top$.
\end{assumption} 

In this assumption, $H$ denotes the horizon length. In the experiments, we use $H=96$, corresponding to 24 hours with a discretization timestep of 15 minutes.

\begin{definition}[Relative Consumption Request]
\label{def:rel_req}
Given a baseline power $p_{t}^{\text{b}}\in\mathbb{R}$, we define a relative consumption request $r_t\in\mathbb{R}$ such that the desired total power at time $t$ is $ p_t=p_{t}^{\text{b}}+r_t$.
\end{definition}

\red{Together with the assumption on the controller, we assume that the equipment of the assets is appropriately sized, thus allowing the asset to return to baseline behavior after relative consumption requests, and having stable behavior. In the case of buildings with HPs, which are used in the simulation examples, this means that the HPs can provide enough heating to let the buildings return to their nominal setpoints after providing flexibility.}

Combining the dynamics \eqref{eq:dyngen}, the constraints \eqref{eq:consbdstategen} and \eqref{eq:conspw}, and relative consumption requests with respect to a baseline consumption as in Definition~\ref{def:rel_req}, we can describe the consumption flexibility of the considered assets in the following way.
\begin{definition}[Flexibility envelope]
\label{def:flex_env}
    Let $\bm{x}_0$ be the starting state of an asset with dynamics described by $h(\cdot)$, $\bm{p}_{0:H-1}^{\text{b}} = [p_{0}^{\text{b}}, \dots, p_{H-1}^{\text{b}}]^\top$ a trajectory of nominal power inputs, and $\bm{e}_{0:H-1}=[\bm{e}_0, \dots, \bm{e}_{H-1}]$ and $\bm{\omega}_{0:H-1}=[\omega_0, \dots, \omega_{H-1}]$ trajectories of external conditions and unmeasured disturbances. With the resulting baseline state trajectory $\bm{x}_{0:H-1}^{\text{b}} = [\bm{x}_{0}^{\text{b}}, \dots, \bm{x}_{H-1}^{\text{b}}]$, recursively defined by $\bm{x}_{t+1}^{\text{b}} = h(\bm{x}_{t}^{\text{b}}, p_{t}^{\text{b}}, \bm{e}_t, \omega_t)$ and $\bm{x}_{0}^{\text{b}}=\bm{x}_0$, we define the relative, fixed-time flexibility envelope $\bm{R}^k_{0:H-1} = [\ubar{\bm{R}}^{k}_{0:H-1}, \bar{\bm{R}}^{k}_{0:H-1}] \in \mathbb{R}^{2\times H}$ with request duration $k$ through
    \begin{align}
        \ubar{\bm{R}}^{k}_{t} = \min& \quad r \label{eq:ftflexminrel}\\
        \bar{\bm{R}}^{k}_{t} = \max& \quad r \label{eq:ftflexmaxrel}
    \end{align}
    both subject to the constraints
    \begin{align*}&\quad \bm{x}_{l+1} = h(\bm{x}_{l}, p_{l}^{\text{b}} + r, \bm{e}_l, \omega_l), \quad  l=t,\dots, t+k-1\nonumber \\
    &\quad \bm{x}_t = \bm{x}_{t}^{\text{b}}  \nonumber\\
    &\quad \ubar{\bm{x}} \leq \bm{x}_{l} \leq \bar{\bm{x}}, \quad l=t,\dots, t+k\nonumber \\
    & \quad \ubar{p} \leq p_{l}^{\text{b}} + r \leq \bar{p}, \quad l=t,\dots, t+k-1.\nonumber
    \end{align*}
    \iffalse
    \begin{align}
        \ubar{\bm{R}}^{k}_{t} = \min& \quad r \nonumber\\
    \text{s.t. } &\quad \bm{x}_{l+1} = h(\bm{x}_{l}, p_{l}^{\text{b}} + r, \bm{e}_l, \omega_l), \nonumber \\ & \qquad \qquad \qquad \qquad l=t,\dots, t+k-1\nonumber \\
    &\quad \bm{x}_t = \bm{x}_{t}^{\text{b}} \label{eq:ftflexminrel}\\
    &\quad \ubar{\bm{x}} \leq \bm{x}_{l} \leq \bar{\bm{x}}, \quad l=t,\dots, t+k\nonumber \\
    & \quad \ubar{p} \leq p_{l}^{\text{b}} + r \leq \bar{p}, \quad l=t,\dots, t+k-1\nonumber
    \end{align}
    \begin{align}
        \bar{\bm{R}}^{k}_{t} = \max& \quad r \nonumber\\
    \text{s.t. } &\quad \bm{x}_{l+1} = h(\bm{x}_{l}, p_{l}^{\text{b}} + r, \bm{e}_l, \omega_l), \nonumber \\ & \qquad \qquad \qquad \qquad l=t,\dots, t+k-1\nonumber \\
    &\quad \bm{x}_t = \bm{x}_{t}^{\text{b}} \label{eq:ftflexmaxrel}\\
    &\quad \ubar{\bm{x}} \leq \bm{x}_{l} \leq \bar{\bm{x}}, \quad l=t,\dots, t+k \nonumber\\
    & \quad \ubar{p} \leq p_{l}^{\text{b}} + r \leq \bar{p}, \quad l=t,\dots, t+k-1\nonumber
    \end{align}
    \fi
\end{definition}

\begin{remark}
    This definition of flexibility envelopes describes the minimum and maximum power levels by which the baseline consumption can be changed, without violating the constraints for a number of $k$ timesteps. These exact flexibility envelopes can not be determined in practice due to modeling errors, forecasting errors (of the baseline consumption and the external conditions), and incomplete information (of the unknown disturbances). However, having a model of the dynamics, the envelopes can be predicted for different systems, e.g., for batteries by using linear state equations. A way to predict these envelopes for thermal assets, while being risk-aware, is summarized in the next section.
\end{remark}

This way of describing flexibility is similar to the power shifting capability described in \cite{reyndersphd}, but uses a fixed time duration. Thus, it also relates to methodology F in \cite{REYNDERS2018review}.

\subsection{Virtual storage modeling of thermal assets}
\label{subsec:batmod}
This section provides a short overview of the data-driven virtual \red{storage} modeling introduced in \cite{batmodel}. For a more in-depth discussion of the setting and assumptions, we refer to the corresponding paper.

To characterize the dynamic behavior of a building, or a general thermal asset, with a virtual \red{storage} model, we introduce a scalar state $s_t\in \mathbb{R}$. This \red{can be seen as an equivalent of the state of charge in batteries and }represents the state of \red{the building} at time $t$ with respect to its thermal mass. Due to this specific type of one-dimensional lumped state, we denote it by $s_t$ instead of $\bm{x}_t$. In this paper, we use the definition \begin{equation}
    s_t := \frac{\ubar{\Delta}_t}{\ubar{\Delta}_t + \bar{\Delta}_t}
\end{equation}
where $\ubar{\Delta}_t$ is the maximum runtime of the equipment at minimum power, and $\bar{\Delta}_t$ is the maximum runtime of the equipment at maximum power, without violating constraints. $\ubar{\Delta}$ and $\bar{\Delta}_t$ were previously introduced in \cite{gasser_predictive_2021}. By definition, we have that $s_t\in [0,1]$, with $s_t=0$ indicating that no thermal energy can be extracted without violating constraints, and $s_t=1$ indicating that no energy can be injected without violating constraints. This state is in general stochastic, due to disturbances.

Furthermore, we consider controlled systems, meaning that the assets are already equipped with a controller that satisfies constraints during normal operation. When receiving relative consumption requests, the controller follows these requests, which leads to a switched system behavior, distinguishing between phases without requests and phases with requests. Under these assumptions, we approximate the state evolution in response to relative consumption requests from the baseline consumption, by the following difference equation
\begin{equation}
    \hat{s}_{t+1} ={ } \hat{s}_t + a^+ r_t^+ + a^- r_t^- + b_f (f(\bm{e}_t) - \hat{s}_t) \chi_{r_t}  + f(\bm{e}_{t+1})-f(\bm{e}_t) \label{eq:stateevol}
\end{equation}
where $r_t^+=\max(r_t, 0)$ and $r_t^-=\min(r_t, 0)$ denote the positive and negative part of the relative consumption request, $\bm{e}_t$ the external conditions used for predicting the baseline consumption, and $\chi_r = \begin{cases}1, \text{ if } r = 0 \\ 0, \text{ if } r \neq 0 \end{cases}$ is the indicator function for request-free periods. The baseline state evolution is assumed to be solely determined from the external conditions $\bm{e}_t$ and denoted by the function $f:\mathbb{R}^m \to \mathbb{R}$. \red{Arbitrary function approximation techniques can be used to learn the function $f$, in the experiments we use a kernel ridge regression model.}

Regarding the \red{virtual} \red{storage} model parameters $a^+, a^-$, and $b_f$, we make the following assumption.
\begin{assumption}
    In \eqref{eq:stateevol}, we assume that
    \begin{enumerate}
        \item $b_f\in\mathbb{R}$ is a constant, 
        \item $a^{+}$ and $a^{-}$ are real-valued random variables on a finite probability space. 
    \end{enumerate}
\end{assumption}
\begin{remark}
    The derivation of the virtual \red{storage} model formulation can be found in \cite{batmodel}. The $a^+$ and $a^-$ terms indicate the linearization of the state change when receiving consumption requests, the $b_f$ term captures the controller behavior of bringing the state back to its nominal value after request periods, and the difference of the nominal state predictions gives the state change incurred through a change in the external conditions. Assuming $a^+, a^-$ to be random variables captures the potential stochasticity of the state evolution due to unknown and unpredictable influences. \red{This includes disturbances through internal gains and inaccuracies in, e.g., predictions of external variables. Like this, no fixed assumption on the source of the disturbance is made, but only its effects in the data are taken into account}. The finiteness assumption is due to the data-driven nature of the sample identification and the inherent constraints on data collection (i.e. time-limited identification phase with restrictions on the extent of the consumption requests), where we denote the set of samples for $a^+$ by $\mathcal{P}^+$, the set of samples for $a^-$ by $\mathcal{P}^-$.%, and the set of all $j$-point averages of parameter tuples as $\mathcal{P}_j = \{\frac{1}{j} \sum_{i=1}^j\bm{a}_i: \bm{a}_i\in\mathcal{P}^+\times\mathcal{P}^-, \bm{a}_l\neq\bm{a}_k\text{ for }l\neq k\}$. The number of parameter tuples is given by $N$.
    \red{The model identification is outlined in Appendix~\ref{app:modelid}.}
\end{remark}

\subsection{Uncertainty-aware flexibility envelopes}

Using the \red{virtual} \red{storage} model \eqref{eq:stateevol} as a description of the dynamics and forecasts for the unknown quantities, we can predict the available consumption flexibility with flexibility envelopes according to Definition~\ref{def:flex_env}, while incorporating a measure of uncertainty. This is done by considering probabilistic constraints with respect to the parameters $a^+$, $a^-$ and reformulating them with a robust uncertainty set using risk measures, as presented in \cite{batmodel}. 

The steps to arrive at the definition of the uncertainty-aware flexibility envelopes $\bm{R}^{\alpha, k}_{0:H-1}=[\ubar{\bm{R}}^{\alpha, k}_{0:H-1}, \bar{\bm{R}}^{\alpha, k}_{0:H-1}]$ with uncertainty level $\alpha$ are given in \ref{app:uaflex}. Here, we restrict ourselves to stating the form derived for the efficient computation of these envelopes:

\begin{lemma}
\label{cor:comftflex}
    Let $a^-_{\max, j} := \max_{(a^+, a^-)\in\mathcal{P}_j} a^-$, $a^+_{\max, j} := \max_{(a^+, a^-)\in\mathcal{P}_j} a^+$, and $\alpha = \frac{j}{N}$. Then
    \begin{align}
        \ubar{\bm{R}}^{\alpha, k}_t &= \max \left\{\max_{l=1,\dots, k} \frac{-f(\bm{e}_{t+l})}{a^-_{\max, j} l} , \max_{l=0,\dots, k-1}\ubar{p}-p_{t+l}^{\text{b}}\right\}, \\
        \bar{\bm{R}}^{\alpha, k}_t &= \min \left\{\min_{l=1,\dots, k} \frac{1-f(\bm{e}_{t+l})}{a^+_{\max, j} l} , \min_{l=0,\dots, k-1}\bar{p}-p_{t+l}^{\text{b}}\right\}.
    \end{align}
\end{lemma}
\begin{proof}
 See \ref{app_proof}.
\end{proof}

\begin{figure}
    \centering
    \input{flex_env}
    \caption{Power increase (green dotted line) and decrease (red dotted line) potential for a duration of 3h with respect to a baseline consumption (blue line), computed with $\alpha = 1$.}
    \label{fig:flex_env_ex}
\end{figure}
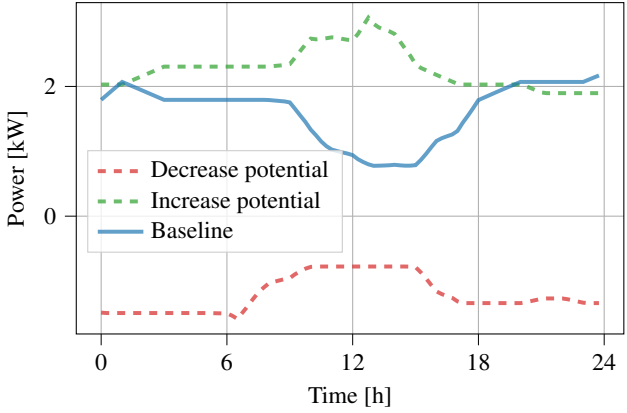

The choice of $j$, and therefore $\alpha$, determines the conservativeness of the predicted available consumption flexibility. An example of the flexibility envelope with $\alpha=1$ is given in Figure~\ref{fig:flex_env_ex}.

\section{Scheduling and dispatch}
\label{sec:sched} 
In this section, we present the main contribution of this work. We use the uncertainty-aware flexibility envelopes described in the previous section to formulate a scheduling problem at the aggregator level and to fulfill an aggregated relative consumption request with multiple assets. This scheduling is determined in advance, and the dispatch of the individual requests to the active assets is carried out with a heuristic algorithm upon receiving each individual request. Through the use of uncertainty-aware flexibility envelopes, the scheduling problem inherits its uncertainty-aware nature, but it also works with general flexibility envelopes of the form presented in Section~\ref{sec:flex_est}.

After specifying a general scheduling problem for request following, we introduce two variations of it, tailored to the objective of fulfilling requests for self-consumption and peak power reduction. One feature of these variations is that they either determine a schedule to track the request if it is fulfillable or propose a new request, that maximizes the flexibility provided.

\subsection{General scheduling problem}
We consider the problem of tracking an aggregated relative consumption request $\bm{r}^{\text{agg}}_{0:H-1}=[r^{\text{agg}}_0, \dots, r^{\text{agg}}_{H-1}]$ for a horizon of $H$ discrete timesteps, by controlling a pool of $M$ assets, where this request is known in advance. A MILP is formulated to determine the activation of consumption flexibility for $M$ flexible assets, where we assume that each asset can be activated only once for a period of $k$ timesteps. The flexibility envelopes of the different assets are given by $\bm{R}^{\alpha, k}_{i, 0:H-1}, i=1,\dots, M$. %Furthermore, we assume in this part that the request is fulfillable, meaning there exists a schedule to cover all requests $r^{\text{agg}}_t$ with a safety margin $\epsilon$.

Different objective functions are possible, here we minimize the number of activated assets to keep flexibility in reserve. The general scheduling problem then looks as follows.

\begin{mini!}
{u_{i,t}}{\sum_{i=1}^M  \sum_{t=0}^{H-1} u_{i,t}}{\label{opt:gensched}}{\label{opt:genschedobj}}
\addConstraint{u_{i,t}\in \{0,1\}, \; i=1,\dots, M, \; t=0,\dots, H-1}{}{\label{opt:genschedc1}}
\addConstraint{\sum_{t=0}^{H-1} u_{i,t}\leq 1, \; i=1,\dots, M}{}{\label{opt:genschedc2}}
\addConstraint{r^{\text{agg}}_t-\sum_{i=1}^M \sum_{l=t_s}^{t}u_{i,l}\ubar{\bm{R}}^{\alpha,k}_{i,l} \geq \epsilon, \; t=0,\dots, H-1}{}{\label{opt:genschedc3}}
\addConstraint{r^{\text{agg}}_t-\sum_{i=1}^M \sum_{l=t_s}^{t}u_{i,l} \bar{\bm{R}}^{\alpha,k}_{i,l}\leq -\epsilon, \; t=0,\dots, H-1}{}{\label{opt:genschedc4}}
\end{mini!}
with $t_s = \max\{t-k-1, 0\}$.
Constraint \eqref{opt:genschedc1} specifies the binary nature of the activation of building $i$ at time $t$, \eqref{opt:genschedc2} states that each asset can only be activated once over the time horizon, and \eqref{opt:genschedc3} and \eqref{opt:genschedc4} describe the covering of the aggregated requests by the flexibility of the activated assets, which stay active for $k$ timesteps.

The result $\bm{U}\in \{0,1\}^{M\times H}$ with $[\bm{U}]_{i,t}=u_{i,t}$ can be interpreted as follows. If $u_{i,t}=1$, then asset $i$ is activated at time $t$ and can provide consumption flexibility within the range $[\ubar{\bm{R}}^{\alpha,k}_{i,t}, \bar{\bm{R}}^{\alpha,k}_{i,t}]$ for $k$ timesteps.

\red{\begin{remark}
    In this formulation, uncertainty is accounted for in the computation of the flexibility envelope. By doing so, the resulting optimization problem corresponds to a standard formulation rather than a robust or stochastic one, rendering it computationally cheaper.
\end{remark}}

Note that the sign of the entries of $\bm{r}^{\text{agg}}_{0:H-1}$ can be arbitrary, so we can respond to purely positive, purely negative, or mixed-sign consumption requests with this approach. Also note that problem~\eqref{opt:gensched} is infeasible if the aggregated request trajectory $\bm{r}^{\text{agg}}_{0:H-1}$ can not be followed with the available flexibility. Therefore, we introduce a scheduling formulation, that determines a committed request trajectory that is fulfillable.

\subsection{Scheduling with request commitment}
To circumvent the case of an infeasible scheduling problem, we reformulate the constraints of \eqref{opt:gensched}, using pointwise scaling factors $d_t\in [0,1], t=0,\dots, H-1$. The constraints look as follows.
\begin{alignat}{1}
   & u_{i,t}\in \{0,1\},\quad i=1,\dots, M,\; t=0,\dots, H-1 \label{eq:commcons1} \\
&\sum_{t=0}^{H-1} u_{i,t}\leq 1,\quad i=1,\dots, M \\
&d_t r^{\text{agg}}_t-\sum_{i=1}^M \sum_{l=t_s}^{t}u_{i,l}\ubar{\bm{R}}^{\alpha,k}_{i,l} \geq \epsilon,\quad t=0,\dots, H-1 \\
&d_tr^{\text{agg}}_t-\sum_{i=1}^M \sum_{l=t_s}^{t}u_{i,l} \bar{\bm{R}}^{\alpha,k}_{i,l}\leq -\epsilon,\quad t=0,\dots, H-1 \\
&0\leq d_t\leq 1,\quad t=0,\dots, H-1 \label{eq:commconslast}
\end{alignat}
with $t_s = \max\{t-k-1, 0\}$.
The new request trajectory $\bm{r}^{\text{comm}}_{0:H-1}=[d_0 r^{\text{agg}}_0, \dots, d_{H-1} r^{\text{agg}}_{H-1}]$ is then referred to as the committed request trajectory. These constraints can in principle be combined with many different objective functions, aiming to maximize the provided flexibility through the committed requests, according to different metrics. In the following, we discuss the cases of self-consumption on a pool level and peak reduction.

\subsection{Scheduling for self-consumption and peak reduction}
To adapt the scheduling problem to the scenarios of self-consumption and peak reduction, we first define these two types of requests, which represent the ideal scenario for the requesting party.

For self-consumption, the goal is to absorb excess power production by increasing the consumption of the flexible assets. This excess production is, e.g., due to high PV power production. Due to this increase of consumption, the request trajectory has only non-negative entries. In this work, we consider self-consumption on a pool level, meaning that we want to absorb the combined production from assets in the pool by the aggregated consumption. Therefore, it looks as follows.
\begin{definition}[Self-consumption request trajectory]
\label{def:reqself}
    Given an aggregated baseline consumption trajectory $\bm{p}^{\text{b},\text{agg}}_{0:H-1}=[p^{\text{b}, \text{agg}}_{0}, \dots, p^{\text{b}, \text{agg}}_{H-1}]\in \mathbb{R}^H$ and an aggregated production trajectory $\bm{g}^{\text{agg}}_{0:H-1}=[g_0^{\text{agg}}, \dots, g^{\text{agg}}_{H-1}]\in \mathbb{R}^H$, the self-consumption request trajectory is given by $\bm{r}^{\text{self}}_{0:H-1}=[r^{\text{self}}_{0}, \dots, r^{\text{self}}_{H-1}]\in \mathbb{R}^H$ with
    \begin{equation}
    \label{eq:selfconreq}
        r^{\text{self}}_{t} := \max \{g^{\text{agg}}_t-p^{\text{b}, \text{agg}}_{t}, 0 \}\quad t=0,\dots,H-1.
    \end{equation}
\end{definition}

This self-consumption request is motivated by definitions like the one given in \cite{LUTHANDER201580} or \cite{resflexrev} and references therein, which is also used as a metric in the evaluation, presented in \ref{subsec:evalmetr}. Tracking this request exactly would lead to a complete self-consumption according to that definition. The scheduling problem to maximize the self-consumption then looks as follows:
\begin{maxi}
{u_{i,t}, d_t}{\sum_{t=0}^{H-1} r^{\text{self}}_t d_t}{\label{opt:selfsched}}{}
\addConstraint{\eqref{eq:commcons1}-\eqref{eq:commconslast}.}{}{}
\end{maxi}

For peak reduction, on the other hand, the requests are non-positive, limiting the overall consumption to a desired level. 
\begin{definition}[Peak reduction request trajectory]
\label{def:reqpeak}
     Given an aggregated baseline consumption trajectory $\bm{p}^{\text{b}, \text{agg}}_{0:H-1}=[p^{\text{b}, \text{agg}}_{0}, \dots, p^{\text{b}, \text{agg}}_{H-1}]\in \mathbb{R}^H$ and a desired peak $c\in \mathbb{R}$, the peak reduction request trajectory $\bm{r}^{\text{peak}}_{0:H-1} = [r^{\text{peak}}_{0}, \dots, r^{\text{peak}}_{H-1}]\in \mathbb{R}^H$ is defined as
     \begin{equation}
        \label{eq:peakredreq}
         r^{\text{peak}}_t := \min \{c - p^{\text{b}, \text{agg}}_{t}, 0\} \quad t=0,\dots,H-1.
     \end{equation}
\end{definition}

This can be interpreted as peak-clipping, already introduced in the 1980's, e.g., in \cite{gellings1985}, and also addressed in \cite{HIRMIZ2019103}. In the scheduling, we want to determine a scaling of the original request that minimizes the new peak. This new peak is denoted by $\rho \in \mathbb{R}$ and lower bounded by the desired peak $c$ if there is at least one non-zero request.
\begin{mini}
{u_{i,t}, d_t, \rho}{\rho}{\label{opt:peaksched}}{}
\addConstraint{\eqref{eq:commcons1}-\eqref{eq:commconslast}}{}{}
\addConstraint{\rho\geq p^{\text{b},\text{agg}}_{t} + d_tr^{\text{peak}}_t}{}{\quad t=0,\dots, H-1}.
\end{mini}

The role of the scaling variable $d_t$ can be understood as follows for the two cases: For the self-consumption case, scaling the incoming request means consuming as much self-produced power as possible without violating the available flexibility ranges given by the flexibility envelopes, which is at a lower level that the overall request. For the peak-reduction case, scaling means reducing the peak as much as possible \red{while fulfilling constraints, thus} achieving a peak consumption that is \red{at least as high as} what was originally requested.

\red{Note that this scaling approach aims at constraint fulfillment and is therefore compatible with arbitrary objectives in the request commitment framework.}

%In the implementation of the peak reduction scheduling problem \eqref{opt:peaksched}, we reformulate the constraint \eqref{opt:peakschedc5} to remove the need for the baseline consumption prediction, since that information is already contained in the request itself, due to Definition~\ref{def:reqpeak}. This is done by minimizing $\tilde{\rho} = \rho - c$ instead, where knowledge of $c$ is not needed to determine the schedule and the request promise since it is a constant. The new constraint is given in the following lemma.
%\begin{lemma}
%    With the change of variable $\tilde{\rho} = \rho - c$, constraint \eqref{opt:peakschedc5} is equivalent to 
%    \begin{equation}
%        r^{\text{peak}}_t \geq d_t r^{\text{peak}}_t - \tilde{\rho}, \quad t=0,\dots,H-1.
%    \end{equation}
%\end{lemma}
%\begin{proof}
%\end{proof}

\subsection{Dispatch algorithm}
Once a schedule $\bm{U}$ has been determined through solving \eqref{opt:gensched}, \eqref{opt:selfsched}, or \eqref{opt:peaksched}, request values still need to be computed for each activated asset. For a committed request $r_t^{\text{comm}}$ at time $t$, this can be done via solving another optimization problem with the following constraints:
\begin{align}
    &\sum_{i=1}^M r_{i,t} = r_t^{\text{comm}} \\
    &\sum_{l=t_s}^t u_{i,l} \ubar{\bm{R}}^{\alpha, k}_{i, l} \leq r_{i,t} \leq \sum_{l=t_s}^t u_{i,l} \bar{\bm{R}}^{\alpha, k}_{i, l} \quad i=1,\dots,M
\end{align}
with $t_s = \max\{t-k-1, 0\}$. Different objectives are again possible in this case. For example, to achieve a balanced activation over all assets, it is possible to minimize $\sum_{i=1}^M r_{i,t}^2$.

Instead of solving an optimization problem, heuristic methods can be used to achieve a fast dispatch. For this, we use the shorthand notation
\begin{equation}
    A_{i,t} = \sum_{l=t_s}^t u_{i,l} \in \{0,1\}
\end{equation}
with $t_s = \max\{t-k-1, 0\}$, to indicate whether asset $i$ is active at time $t$. Then we can formulate the heuristic dispatch through
\begin{equation}
    \label{eq:indivreq}
    r_{i,t} = A_{i,t} \frac{F_i}{\sum_{j=1}^M F_j A_{j,t}} r^{\text{comm}}_t
\end{equation}
where $F_i$ denotes a measure of the available flexibility of asset $i$. 
\iffalse Possible choices of this are the available activated flexibility, by choosing 
\begin{equation}
    F_i = \sum_{t=0}^{H-1} u_{i, t} \bar{\bm{R}}^{\alpha, k}_{i, t}
\end{equation}
or 
\begin{equation}
    F_i = \sum_{t=0}^{H-1} u_{i, t} \ubar{\bm{R}}^{\alpha, k}_{i, t}
\end{equation}
depending on the sign of the request $r^{\text{comm}}_t$. \fi 
A possible choice is to use the average flexibility potential as a proxy, given by
\begin{equation}
\label{eq:flexindheur}
    F_i = \frac{1}{H} \sum_{t=0}^{H-1} (\bar{\bm{R}}^{\alpha, k}_{i, t}- \ubar{\bm{R}}^{\alpha, k}_{i, t}).
\end{equation}
\eqref{eq:flexindheur} is used for computational efficiency in the experiments.

\red{\begin{remark}
    Following consumption requests during certain periods can introduce effects on the consumption after those periods. This phenomenon, known as rebound effect, depends on the time it takes the controller to steer the asset back to its nominal state and thus consuming more or less than its baseline during that time. In the experiments, a slow rebound after high consumption requests can be observed.
\end{remark}}

\section{Simulation setup}
\label{sec:exp}
The proposed approach is tested in simulation, using a pool of buildings from the Python library Energym \cite{scharnhorst_energym}. The scheduling problems are formulated with the linear programming toolkit PuLP \cite{Mitchell2011PuLPA} and solved with CBC \cite{john_forrest_2023_cbc} or Gurobi \cite{gurobi}.

\subsection{Building models and requests}
\label{subsec:bdmodreq}
We use the SimpleHouseRad-v0 model from Energym as a flexible asset. This represents a lightweight single-family house, modeled as a single zone, equipped with a HP. We sample the building parameters of thermal capacity, thermal conductance, and nominal Coefficient of Performance (COP) of the HP uniformly at random from pre-specified intervals, furthermore, we scale the maximum HP power according to the sampled thermal conductance. This is done to ensure slightly varying characteristics in the overall pool of buildings. \red{The intervals for sampling are as follows: thermal capacity between 2.77 and 11.11 kWh/°C, thermal conductance between 0.2 and 0.5 kW/°C, HP maximum power between 2 and 5 kW, and HP nominal COP between 4 and 5.}

The HP power fraction is controlled by a PID controller with a control timestep of 5 minutes. A temperature setpoint of 21 °C is followed by the PID controller, and we set the acceptable temperature to the range $[19, 24]$ °C. Flexibility predictions and requests are on the other hand sent with a 15-minute timestep. \red{The complete procedure is summarized in Algorithm~\ref{alg:VERM2}.}

Experiments of the coordination of 100 to 500 buildings for the scenarios of self-consumption and peak reduction are performed, as well as scalability experiments for solving the scheduling problem with up to 2000 buildings. \red{The used building parameters for all 500 buildings and the flexibility envelopes used for the scalability experiments can be found in the GitHub repository \url{https://github.com/psh987/flex_scheduling}.}

Self-consumption requests are generated as follows. An aggregated production curve is computed, using the PVSystem class of the Python library pvlib \cite{Holmgren2018} as a single system, with a capacity scaled to the overall pool of buildings. For this, perfect forecasts of the irradiance and temperature are used, provided by Energym. The request is then given by the difference of the production and the aggregated baseline consumption prediction, as stated in Definition~\ref{def:reqself}.

Peak reduction requests are based on real consumption data from the canton of Neuchâtel, Switzerland, provided by the grid operator Swissgrid \cite{swissgrid_data_2022}. This consumption data is scaled down to match the magnitude of the consumption of the pool of buildings. The overall consumption is assumed to be the sum of the baseline consumption of the pool of buildings and a non-shiftable baseline. Thus, changing the consumption patterns of the buildings is the only way to achieve a peak reduction. We define a desired new peak $c$, in the experiments chosen as $1$ to $1.1$ times the average consumption for that particular day, and then compute the request based on Definition~\ref{def:reqpeak} with respect to the scaled real consumption.

\begin{algorithm*}
\red{
\caption{Complete procedure of the approach}\label{alg:VERM2}
\begin{algorithmic}[1]
\State \textbf{Input:} Asset data, daily weather forecasts
\State \textbf{Main Execution:}
\State Train the virtual \red{storage} models for all flexible assets (according to Equation~\ref{eq:stateevol})

\For{each day}
    \State \textbf{Individual Assets:}
    \State Compute their day ahead flexibility envelope (according to Lemma~\ref{cor:comftflex})

    \State \textbf{Aggregator:}
    \State Receive flexibility envelopes from all flexible assets
    \State Receive aggregate relative consumption request (e.g. Equations \ref{eq:selfconreq}, \ref{eq:peakredreq})
    \State Solve the day ahead scheduling problem (according to Equations \ref{eq:commcons1}-\ref{eq:commconslast})
    \For{every 15 minutes}
        \State Compute and apply the individual consumption request per asset (according to Equation \ref{eq:indivreq})
    \EndFor
\EndFor

\end{algorithmic}}
\end{algorithm*}

\subsection{Baseline approach}
\label{subsec:greedy}
We compare our approach to a simple greedy strategy for responding to requests. For this, assets are grouped into three sets: available assets, active assets, and inactive assets. Upon receiving a consumption request, it is checked if this request is fulfillable with the available flexibility of all buildings in the active group. If it is not fulfillable, buildings from the available group are activated until the request is either fulfillable or no buildings are left in the available group. This activation is done at random. Buildings stay active for at most $k$ timesteps, after that they are set as inactive. Violating the comfort bounds also leads to a building being set as inactive.

%To mitigate the rebound of the inactive assets and to guarantee comparability in the experiments, the approach outlined in Section~\ref{subsec:rebound} is used. 

\subsection{Evaluation metrics}
\label{subsec:evalmetr}

In the experiments, we distinguish between performance metrics and comfort metrics. As performance metrics for the peak reduction case, we use the absolute peak power reduction, defined as follows.
\begin{definition}
Given the aggregated baseline consumption $\bm{p}^{\text{b}, \text{agg}}_{0:H-1}$ and the aggregated actual consumption $\bm{p}^{\text{agg}}_{0:H-1}$, we define the absolute peak power reduction (absolute PPR) as
    \begin{equation}
    \Delta P_a = \max(\bm{p}^{\text{b}, \text{agg}}_{0:H-1}) - \max(\bm{p}^{\text{agg}}_{0:H-1}).
    \label{eq:abspeakpowred}
\end{equation}
%and the peak power reduction fraction as
%\begin{equation}
%    \Delta P_r = 1-\frac{\max(\bm{p}^{\text{agg}}_{\text{a}, 0:H-1})}{\max(\bm{p}^{\text{agg}}_{\text{n}, 0:H-1})}.
%    \label{eq:peakpowredfrac}
%\end{equation}
\end{definition}

For the case of self-consumption, we consider the metric of self-consumed power fraction, defined as follows.
\begin{definition}
    Given the aggregated production $\bm g^{\text{agg}}_{0:H-1}$, the production sum $g_{\text{sum}} = \sum_{t=0}^{H-1}g^{\text{agg}}_t$, and the aggregated actual consumption $\bm{p}^{\text{agg}}_{0:H-1}$. Then the self-consumed power fraction is given by
    \begin{equation}
        \Delta S_r = \frac{g_{\text{sum}} - \sum_{t=0}^{H-1} \max\{g^{\text{agg}}_t-p^{\text{agg}}_t, 0\}}{g_{\text{sum}}}.
        \label{eq:selfconrel}
    \end{equation}
\end{definition}
This metric specifies how much of the overall production was directly consumed. These performance metrics are, e.g., presented in \cite{resflexrev}.

To measure comfort, we specify the percentage of temperature-bound violations for a specific acceptable temperature interval. This is defined as follows.
\begin{definition}
    Given a temperature trajectory $\bm T_{0:H-1} = [T_0, \dots, T_{H-1}]$ and temperature bounds $\ubar{T}, \bar{T}$, we define the percentage of temperature bound violations as 
    \begin{equation}
        \Delta T_r = \frac{\sum_{t=0}^{H-1}I_{\ubar{T}}^{\bar{T}}(T_t)}{H} 100\%
        \label{eq:percviol}
    \end{equation}
    with $I_a^b(c)=\begin{cases}
        0\text{, if }c\in [a,b] \\
        1\text{, if }c\not\in [a,b]
    \end{cases}$.
\end{definition}
This metric is similar to the prediction interval coverage percentage, frequently used in statistical forecasting \cite{GONZALEZSOPENA2021110515}.

For the scalability experiments, we determine the solving times of the scheduling problem up to a predefined gap and report the average time for multiple runs as a metric for scalability.

\section{Results}
\label{sec:res}
In this section, we collect the results of the different simulation experiments. The experiments include the application of our approach to the scenarios of self-consumption and peak reduction, a comparison with the baseline approach explained in Section~\ref{subsec:greedy}, a variation of the allowed activation timesteps $k$ for a fixed number of buildings, a variation of the number of buildings for a fixed request, and the scalability experiments. In the comparisons, we also consider the effect of different uncertainty parameters $\alpha$ in the computation of the individual flexibility envelopes (see Lemma~\ref{cor:comftflex} for further details on $\alpha$).

For all building models, we collect data from the first 21 days of the year to fit the virtual \red{storage} models described in Section~\ref{subsec:batmod}. For results on the flexibility prediction accuracy, we refer to \cite{batmodel}. The test period covers the 50 following days. To evaluate the results for peak reduction, the metric in \eqref{eq:abspeakpowred} is computed for each day and then averaged over the 50 days. For the self-consumed power fraction and the percentage of temperature violations, \eqref{eq:selfconrel} and \eqref{eq:percviol} are computed over the whole 50 days. The installed controllers of the individual buildings are assumed to be rebound-aware, therefore limiting their deviation from the baseline consumption after request periods. In the experiments, this is expressed as a maximum allowed deviation from the baseline of 20\%. %To mitigate rebound effects in all the tested scenarios, we limit the deviation from the baseline consumption for the period after the request period.

\subsection{Self-consumption experiments}
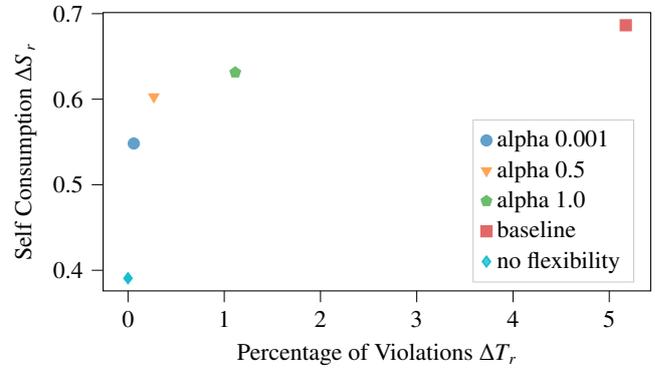
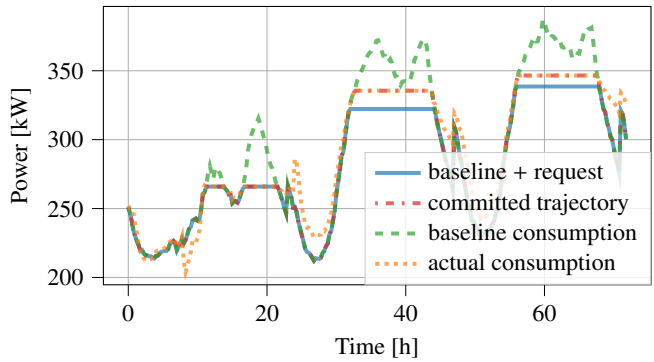
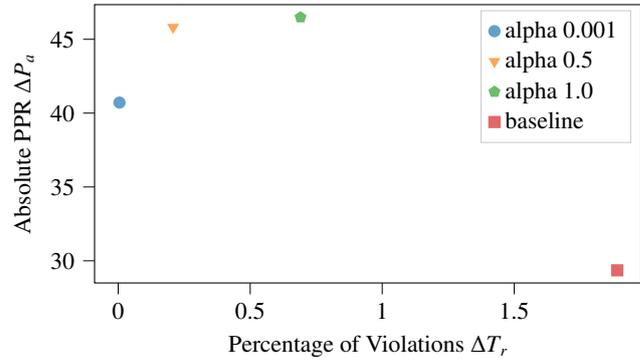
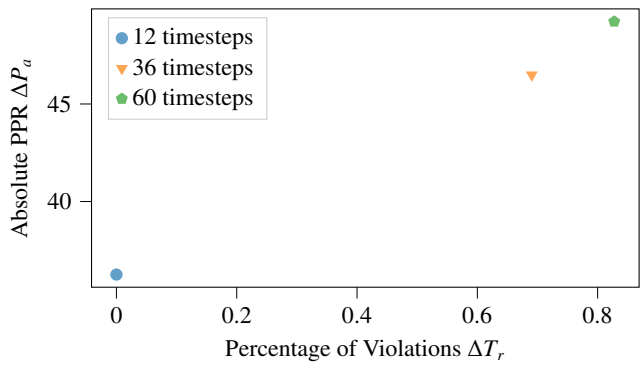
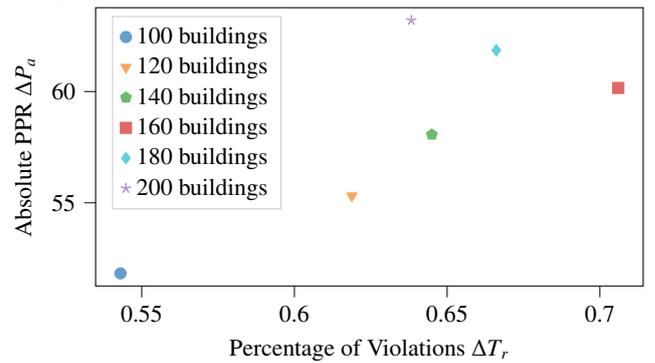
\begin{figure*}
    \centering
    \begin{subfigure}[b]{0.49\textwidth}
        \input{self_con_3days}
        \caption{Self-consumption request following for the first three days of the test period with $\alpha=1$.}
        \label{fig:self_exp_follow}
    \end{subfigure}
    \hfill
    \begin{subfigure}[b]{0.49\textwidth}
        \input{self_con_eval}
        \caption{Cumulated metrics for the test period with varying $\alpha$ values and comparison to the baseline and nominal controller operation.}
        \label{fig:self_exp_metr}
    \end{subfigure}
    \begin{subfigure}[b]{0.48\textwidth}
        \input{peak_red_3days}
        \caption{Peak reduction request following for the first three days of the test period with $\alpha=1$.}
        \label{fig:peak_exp_follow}
    \end{subfigure}
    \hfill
    \begin{subfigure}[b]{0.48\textwidth}
        \input{peak_red_eval}
        \caption{Averaged metrics for the test period with varying $\alpha$ values and comparison to the baseline.}
        \label{fig:peak_exp_metr}
    \end{subfigure}
    \begin{subfigure}[b]{0.49\textwidth}
        \input{peak_red_var_k}
        \caption{Averaged metrics for the test period with $\alpha=1$ and varying availability times.}
        \label{fig:peak_exp_vartime}
    \end{subfigure}
    \hfill
    \begin{subfigure}[b]{0.49\textwidth}
        \input{peak_red_var_bd}
        \caption{Averaged metrics for the test period with $\alpha=1$, a fixed request, and a varying number of buildings.}
        \label{fig:peak_exp_varscale}
    \end{subfigure}
    \caption{Results of the self-consumption (a, b) and peak reduction (c, d) experiments with 100 buildings and a maximum activation time of 3h per building and impact of varying parameters in the peak reduction experiments with $\alpha=1$ (e, f).}
    \label{fig:self_exp}
\end{figure*}

The results of running the self-consumption experiments for the 50 test days are displayed in Figure~\ref{fig:self_exp}. Figure~\ref{fig:self_exp_follow} shows the first three days of the test period with the baseline consumption forecast given in green, the actual consumption in red, the baseline plus relative request in blue, and the baseline plus committed request in orange. Due to the perfect tracking during the request periods, the orange line is covered by the red one. The first day shows a large request that is not fulfillable with the estimated available flexibility and therefore leads to a lower committed request. On days two and three, the requests are fulfillable and thus tracked exactly. 

Figure~\ref{fig:self_exp_metr} shows the self-consumed power fraction and the percentage of temperature-bound violations for $\alpha$ values of $0.001$, $0.5$, and $1$, and compares them to the baseline approach and the nominal controller operation without receiving flexibility requests. As expected, a lower $\alpha$ leads to more conservative predictions of the available flexibility, and therefore both to a lower self-consumption $\Delta S_r$ and a lower percentage of violations $\Delta T_r$, with about $0.548$ and $0.06\%$ respectively. With increasing $\alpha$, both of the metrics increase as well, up to a $\Delta S_r$ of about $0.631$ and a $\Delta T_r$ of about $1.11\%$ for $\alpha=1$. The baseline approach achieves the highest self-consumption with $\Delta S_r \approx 0.686$, but also a high percentage of violations with $\Delta T_r \approx 5.17\%$. In comparison, the nominal controller operation does not result in any temperature-bound violations and has a self-consumption of $\Delta S_r \approx 0.391$.

We also run the experiments for 500 buildings with our approach and $\alpha=1$. Since the requests are scaled to the number of buildings, a comparable result of $\Delta S_r \approx 0.629$ and $\Delta T_r \approx 1.17\%$ is achieved.

\subsection{Peak reduction experiments}
Figure~\ref{fig:self_exp} also presents the results of the peak reduction experiments for the 50 test days. The results for the first three days with $\alpha=1$ are shown in Figure~\ref{fig:peak_exp_follow}. In this example, the request of the first day is fulfillable and tracked accordingly. For days two and three, the available flexibility is not sufficient to decrease the overall consumption to the desired level (blue line), instead, a smaller decrease is committed to and followed (orange line, covered by red line).

As in the self-consumption case, a smaller $\alpha$ leads to a lower absolute peak power reduction and a lower percentage of violations ($\Delta P_a \approx 40.71$kW, $\Delta T_r \approx 0.005\%$ for $\alpha=0.001$), whereas a higher $\alpha$ leads to better performance, but higher violations ($\Delta P_a \approx 46.49$kW, $\Delta T_r \approx 0.691\%$ for $\alpha=1$). An intermediate $\alpha$ seems to deliver a good tradeoff between performance and violations for both self-consumption and peak reduction. The baseline approach performs less well in the peak reduction case, having both the lowest absolute peak power reduction of about $29.36$kW and the highest percentage of violations of about $1.89\%$.

In the corresponding experiments with 500 buildings and $\alpha=1$, a peak power reduction of about $229.80$kW is achieved with about $0.608\%$ bound violations.
\iffalse
\begin{figure*}
    \begin{subfigure}[b]{0.48\textwidth}
        %\includegraphics[width=\textwidth]{figs/simflex_bd_100_peak_days_50_alpha_1.0_vartol_True_3days.pdf}
        \input{peak_red_3days}
        \caption{Request following for the first three days of the test period with $\alpha=1$.}
        \label{fig:peak_exp_follow}
    \end{subfigure}
    \hfill
    \begin{subfigure}[b]{0.48\textwidth}
        %\includegraphics[width=\textwidth]{figs/evaluation_100_peak_50_reg_exp.pdf}
        \input{peak_red_eval}
        \caption{Averaged metrics for the test period with varying $\alpha$ values and comparison to the baseline.}
        \label{fig:peak_exp_metr}
    \end{subfigure}
    \caption{Results of the peak reduction experiments with 100 buildings and a maximum activation time of 3h per building.}
    \label{fig:peak_exp}
\end{figure*}
\fi
We also run experiments to quantify the change in metrics with a varying number of available timesteps $k$. Similarly, we run experiments to quantify how the number of buildings influences the resulting metrics.

\begin{figure*}[ht]
    \centering
    \input{sol_time_res}
    \caption{Average solving times of the peak reduction scheduling problem for varying numbers of assets in three request cases.}
    \label{fig:scal}
\end{figure*}
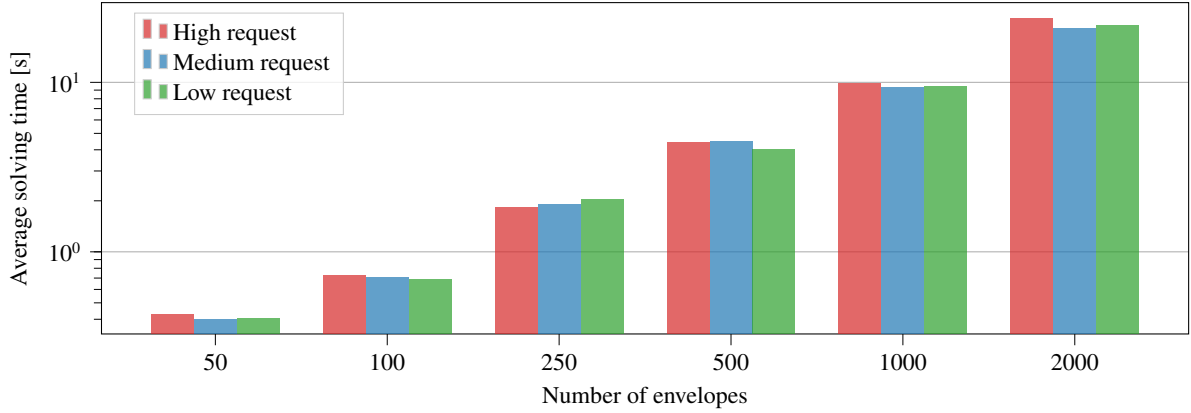
For the first part, we run the experiments for 100 buildings and compute flexibility envelopes for $k=12$ (i.e., 1 hour availability), $k=36$ (i.e., 3 hours availability), and $k=60$ (i.e., 5 hours availability) with $\alpha=1$. The results are shown in Figure~\ref{fig:peak_exp_vartime}. A short availability time leads to no temperature violations in this case, but also a lower absolute peak power reduction of about $36.26$kW. With increasing activation duration, both of these metrics increase to about $0.828\%$ of violations and $\Delta P_a$ of $49.23$kW. This might be due to either a duration of 1 hour not being enough for the building to saturate its temperature bounds, or to a longer prediction horizon leading to an accumulation of errors, and therefore an overestimation of the available flexibility.

For the second part, we generate requests according to the description in Section~\ref{subsec:bdmodreq}, for a pool of 120 buildings. The request following is then attempted with 100 to 200 buildings. The results are shown in Figure~\ref{fig:peak_exp_varscale}. As expected, having more buildings increases the capabilities in peak reduction, which can be observed with an absolute peak power reduction of $51.83$kW for 100 buildings to a $\Delta P_a$ of $63.20$kW for 200 buildings. Interestingly, no clear trend is visible in the percentage of violations, being in the range of $0.543\%$ to $0.706\%$.

\iffalse
\begin{figure*}
    \centering
    \begin{subfigure}[b]{0.49\textwidth}
        %\includegraphics[width=\textwidth]{figs/evaluation_100_peak_50_var_k.pdf}
        \input{peak_red_var_k}
        \caption{Averaged metrics for the test period with $\alpha=1$ and varying availability times.}
        \label{fig:peak_exp_vartime}
    \end{subfigure}
    \hfill
    \begin{subfigure}[b]{0.49\textwidth}
        %\includegraphics[width=\textwidth]{figs/evaluation_200_peak_50_var_bd.pdf}
        \input{peak_red_var_bd}
        \caption{Averaged metrics for the test period with $\alpha=1$, a fixed request, and a varying number of buildings.}
        \label{fig:peak_exp_varscale}
    \end{subfigure}
    \caption{Impact of varying parameters in the peak reduction experiments with $\alpha=1$.}
    \label{fig:peak_exp_var}
\end{figure*}
\fi

\subsection{Scalability experiments}
To test the scalability of our approach to the number of included assets, we measure the wall-clock solving time of the scheduling problem with different numbers of flexibility envelopes. The flexibility envelopes are chosen at random from a set of 25000 envelopes, generated from the operation of 500 buildings during 50 days. The request trajectory is generated for a random day of the year, distinguishing between a high request (new desired peak $c$ as the average consumption of that day), a medium request ($1.05$ times the average), and a low request ($1.1$ times the average) scenario. This is done for 50, 100, 250, 500, 1000, and 2000 envelopes, and the solving times are averaged over 20 runs. The problem is solved up to an absolute gap of $0.01c$, using Gurobi \cite{gurobi} as a solver. These experiments were run on a laptop with Intel i7-8565U processor running at 1.8 GHz and 16 GB of RAM. The results of the experiments are displayed in Figure~\ref{fig:scal}.

The solving times scale approximately linearly in the number of considered envelopes, the peak being reached at an average of 24 seconds for the high request scenario for 2000 envelopes, and the minimum at an average of about 0.4 seconds for the medium request scenario for 50 envelopes. When repeating the experiments with a fixed absolute gap of 1, meaning that the promised peak would be at most 1kW away from the optimally achievable one, the results lie in a comparable range. Considering that solving this problem would usually be done once a day for scheduling the activation of assets, these runtimes suggest the feasibility of our approach for an even larger number of assets. Also, depending on the exact time requirements, the approach could be used in a receding horizon framework for a medium to high number of assets.

\section{Conclusion}
\label{sec:con}
Efficient usage of consumption flexibility of buildings and other flexible assets for grid balancing is becoming increasingly important with growing shares of renewable energy sources and increasing electrification of energy services. This work proposes a coordination method for multiple flexible assets, by scheduling them according to an external flexibility request that reflects the needs of an upper-layer control entity, such as an aggregator or a grid operator. The schedule is determined by solving a mixed-integer linear program, where we propose specific forms of this optimization problem for the scenarios of self-consumption and peak reduction. A feature of these scheduling problems is that they also determine the maximum fulfillable request, in case the original request is not fulfillable with the available flexibility. A general characterization of flexibility via flexibility envelopes is used, and a way to use uncertainty-aware flexibility envelope predictions for the individual assets with a freely selectable risk parameter is presented. Heuristic dispatch and rebound-damping strategies are proposed to deal with incoming requests according to the schedule. The approach, centered on a maximum fixed period of activation of a few hours per asset over the horizon, can be used both in implicit (manual control changes by prosumers) and explicit (automatic control changes) DR schemes.

Simulation results showed that the approach is effective in providing flexibility while respecting comfort constraints. Also, the scalability of the scheduling problem to a high number of assets was demonstrated. 

Different ways of extending this work are envisioned. First, an investigation of the impact of heterogeneous assets in the pool is planned. The present experiments use varying building parameters, and forthcoming works will address different types of assets like water tanks or stationary storage systems. Second, an extension of the approach to a receding horizon formulation is of interest. This line of work is facilitated by the possibility to quickly update flexibility estimations using the \red{virtual} \red{storage} models and to solve the scheduling problem in a reasonable time for a medium to high number of assets. \red{Third, a comparative study with more sophisticated scheduling mechanisms like robust or stochastic optimization is needed to prove the effectiveness of the approach and the computational efficiency. }Lastly, we aim to show the efficacy of our approach in a real-world example.

\section*{Acknowledgements}
This project has received funding from the European Union’s Horizon 2020 research and innovation programme under grant agreement No 101033700, CSEM's Data Program, and the Swiss National Science Foundation under the RISK project (Risk Aware Data-Driven Demand Response, grant number 200021 175627).

\appendix
\red{
\section{Virtual storage model identification}
\label{app:modelid}
The identification of the virtual storage model in Equation~\ref{eq:stateevol} happens in two stages: First, the baseline state model $f(e_t)$ is learned from data of nominal operation of the controller. Second, the parameter samples of $a^+$ and $a^-$ are identified from periods with nonzero requests, and $b_f$ is identified from data of periods where the controller steers the state back to its nominal value after a request.

For the second part, consider a  request trajectory that fulfills the following condition for the smallest possible $l\in\mathbb{N}$: $(\bm{r}_{0:l}>\bm{0}$ or $\bm{r}_{0:l}<\bm{0})$ and $(r_{l+1}=0$ or $s_{l+1}=0$ or $s_{l+1}=1)$. Using this, we can determine samples for $a^+, a^-$ by unrolling Equation~\ref{eq:stateevol}. We have
\begin{equation}
    s_l = s_0 + \sum_{i=0}^{l-1} a^{+/-} r_i + f(\bm{e}_{l})-f(\bm{e}_0)
\end{equation}
and therefore
\begin{equation}
    a^{+/-}=(s_{l}-f(\bm{e}_{l})-(s_0-f(\bm{e}_0))) /\sum_{i=0}^{l-1} r_i
\end{equation}
This can be seen as the average state change depending on the average request, accounting for a state change due to a change in external conditions. Samples of $a^+, a^-$ are generated for all request trajectories fulfilling the condition.

To get $b_f$, we consider state trajectories right after periods of nonzero requests. For this, we consider a threshold of deviation of the observed state $s_t$ and the predicted baseline state $f(\bm{e}_t)$, denoted by $\delta$. The resulting condition for these recovery periods is: $\bm{r}_{0:l}=\bm{0}$ and $(r_{l+1}\neq 0 $ or $|s_{l+1}-f(\bm{e}_{l+1})|<\delta)$. Again, using the unrolled state equation, we get candidates for $b_f$ from all recovery periods $j=1, \dots, M$ by considering
\begin{equation}
    s_l = (1-b_f)^l s_0 - (1-b_f)^l f(\bm{e}_0) + f(\bm{e}_l)
\end{equation}
and solving the least-squares problems
\begin{equation}
    b_j= \arg\min_{b} \left( (1-b)^{l} (s_0 -f(\bm{e}_0))+ f(\bm{e}_{l}) -s_{l}\right)^2.
\end{equation}
In the experiments, $b_f$ is chosen as $\max_j b_j$, but other choices are possible.}

\section{Uncertainty-aware flexibility envelopes}
\label{app:uaflex}

We can make the following statement about probabilistic state constraint satisfaction in the state dynamics \eqref{eq:stateevol}:

\begin{lemma}
\label{lem:probrobcons}
    Let the sample sets of $a^+$ and $a^-$ be denoted by $\mathcal{P}^+$ and $\mathcal{P}^-$, $|\mathcal{P}^+ \times \mathcal{P}^-| = N$, and an uncertainty parameter $\alpha$ chosen as $\alpha=\frac{j}{N}$ for a $j\in\{1,\dots, N\}$. With the set of all $j$-point averages of parameter tuples denoted by $\mathcal{P}_j := \{\frac{1}{j} \sum_{i=1}^j\bm{a}_i: \bm{a}_i\in\mathcal{P}^+\times\mathcal{P}^-, \bm{a}_l\neq\bm{a}_k\text{ for }l\neq k\}$, we have that if for a relative consumption request $r_t\in \mathbb{R}$
    \begin{multline}
    \label{eq:robstatecons}
        0\leq  \hat{s}_t + a^+ r_t^+ + a^- r_t^- + b_f (f(\bm{e}_t) - \hat{s}_t) \chi_{r_t}  + \\ f(\bm{e}_{t+1})-f(\bm{e}_t) \leq 1 \quad \forall (a^+, a^-) \in \mathcal{P}_j,
    \end{multline}
    then 
    \begin{multline}
    \label{eq:probstatecons}
        \mathbb{P} \{ 0\leq \hat{s}_t + a^+ r_t^+ + a^- r_t^- + b_f (f(\bm{e}_t) - \hat{s}_t) \chi_{r_t}  +\\ f(\bm{e}_{t+1})-f(\bm{e}_t) \leq 1 \} \geq 1-\alpha.
    \end{multline}
\end{lemma}
\begin{proof}
    See \cite[Sec. IV.A]{batmodel}.
\end{proof}

The constraint in \eqref{eq:robstatecons} is a tightened version of the probabilistic constraint \eqref{eq:probstatecons}. We can make use of these tightened constraints to define a new, uncertainty-aware version of the flexibility envelopes from Definition~\ref{def:flex_env}.

\begin{definition}[Uncertainty-aware flexibility envelope]
    \label{def:uaflexenv}
    Let $\alpha=\frac{j}{N}\in(0,1], j\in\{1,\dots, N\}$ be an uncertainty parameter that can be freely chosen. Assuming the asset is in its nominal state when starting a request period, the uncertainty-aware flexibility envelope is given by $\bm{R}^{\alpha, k}_{0:H-1}=[\ubar{\bm{R}}^{\alpha, k}_{0:H-1}, \bar{\bm{R}}^{\alpha, k}_{0:H-1}]$ where
    %\begin{align}
    %    \ubar{\bm{R}}^{\alpha, k}_{t} = \min& \quad r \label{eq:uaflexenvmin}\\
    %    \bar{\bm{R}}^{\alpha, k}_{t} = \max& \quad r \label{eq:uaflexenvmax}
    %\end{align}
    %both subject to the constraints
    %\begin{align*}
    %\text{s.t. } &\quad \mathbf{0} \leq \begin{pmatrix}
    %1 \\ \vdots \\ k
    %\end{pmatrix} a^- r + \begin{pmatrix}
    % f(\bm{e}_{t+1}) \\ \vdots \\ f(\bm{e}_{t+k})
    %\end{pmatrix}\leq \mathbf{1}, \quad  \forall (a^+, a^-) \in \mathcal{P}_j\nonumber \\
    %& \quad \ubar{p} \leq p_{l+t}^{\text{b}} + r \leq \bar{p}, \quad l=0,\dots, k-1\nonumber
    %\end{align*}

    \begin{align}
        \ubar{\bm{R}}^{\alpha, k}_{t} = \min& \quad r \label{eq:uaflexenvmin}\\
        \text{s.t. } &\quad \mathbf{0} \leq \begin{pmatrix}
    1 \\ \vdots \\ k
    \end{pmatrix} a^- r + \begin{pmatrix}
     f(\bm{e}_{t+1}) \\ \vdots \\ f(\bm{e}_{t+k})
    \end{pmatrix}\leq \mathbf{1}, \quad  \forall (a^+, a^-) \in \mathcal{P}_j\nonumber \\
    & \quad \ubar{p} \leq p_{l+t}^{\text{b}} + r \leq \bar{p}, \quad l=0,\dots, k-1,\nonumber
    \end{align}

    \begin{align}
        \bar{\bm{R}}^{\alpha, k}_{t} = \max& \quad r \label{eq:uaflexenvmax}\\
        \text{s.t. } &\quad \mathbf{0} \leq \begin{pmatrix}
    1 \\ \vdots \\ k
    \end{pmatrix} a^+ r + \begin{pmatrix}
     f(\bm{e}_{t+1}) \\ \vdots \\ f(\bm{e}_{t+k})
    \end{pmatrix}\leq \mathbf{1}, \quad  \forall (a^+, a^-) \in \mathcal{P}_j\nonumber \\
    & \quad \ubar{p} \leq p_{l+t}^{\text{b}} + r \leq \bar{p}, \quad l=0,\dots, k-1.\nonumber
    \end{align}

    \iffalse
    \begin{align}
        \ubar{\bm{R}}^{\alpha, k}_{t} = \min& \quad r \nonumber\\
    \text{s.t. } &\quad \mathbf{0} \leq \begin{pmatrix}
    1 \\ \vdots \\ k
    \end{pmatrix} a^- r + \begin{pmatrix}
     f(\bm{e}_{t+1}) \\ \vdots \\ f(\bm{e}_{t+k})
    \end{pmatrix}\leq \mathbf{1}\label{eq:uaflexenvmin}  \\ &\qquad \qquad \qquad \qquad \qquad \forall (a^+, a^-) \in \mathcal{P}_j\nonumber \\
    & \quad \ubar{p} \leq p_{l+t}^{\text{b}} + r \leq \bar{p}, \quad l=0,\dots, k-1\nonumber
    \end{align}
    \begin{align}
        \bar{\bm{R}}^{\alpha, k}_{t} = \max& \quad r \nonumber\\
    \text{s.t. } &\quad \mathbf{0} \leq \begin{pmatrix}
         1 \\ \vdots \\ k
    \end{pmatrix} a^+ r + \begin{pmatrix}
         f(\bm{e}_{t+1}) \\ \vdots \\ f(\bm{e}_{t+k})
    \end{pmatrix}\leq \mathbf{1} \label{eq:uaflexenvmax}\\ & \qquad \qquad \qquad \qquad \qquad \forall (a^+, a^-) \in \mathcal{P}_j \nonumber \\
    & \quad \ubar{p} \leq p_{l+t}^{\text{b}} + r \leq \bar{p}, \quad l=0,\dots, k-1\nonumber
    \end{align}
    \fi
\end{definition}
An example of a flexibility envelope prediction from a simulated building with HP is given in Figure~\ref{fig:flex_env_ex}. It shows the flexibility potential of this building for activations of up to 3 hours.

\begin{remark}
    \eqref{eq:uaflexenvmin} and \eqref{eq:uaflexenvmax} are determined by plugging the unrolled state equation \eqref{eq:stateevol} into the constraints of \eqref{eq:ftflexminrel} and \eqref{eq:ftflexmaxrel}, while using the result of Lemma~\ref{lem:probrobcons}. It is sufficient to only consider $a^-$ or $a^+$ respectively, because $\ubar{\bm{R}}^{\alpha, k}_{t}$ is non-positive, and $\bar{\bm{R}}^{\alpha, k}_{t}$ is non-negative, due to the assumption on the controller.% The steps to arrive at Definition~\ref{def:uaflexenv} are explained in more detail in \cite{batmodel}. 
\end{remark}

Note that in this formulation, we assume the $a^+, a^-$ to be constant during request periods, but potentially varying between request periods. However, due to the following observations for efficient computation of the flexibility envelopes, it is equivalent to allowing varying values of $a^+, a^-$: first, due to the baseline, i.e., a zero request, being feasible, we only need to consider the lower bounds in \eqref{eq:uaflexenvmin} and the upper bounds in \eqref{eq:uaflexenvmax}. Second, we can restrict ourselves to guaranteeing constraint satisfaction for the extreme points in $\mathcal{P}_j$, to guarantee constraint satisfaction for all points in $\mathcal{P}_j$. 

\section{Proof of Lemma~\ref{cor:comftflex}}
\label{app_proof}

We will only focus on the reformulation of $\ubar{\boldsymbol{R}}^{\alpha, k}_t$, the result for $\bar{\boldsymbol{R}}^{\alpha, k}_t$ follows with the same arguments.
    
Since $\ubar{\boldsymbol{R}}^{\alpha, k}_t\leq 0$ due to the controller assumption, we only need to consider the lower bounds in \eqref{eq:uaflexenvmin}. This leads to
    \begin{align}
    &\quad \begin{pmatrix}
     -f(\boldsymbol{e}_{:t+1})/a^- \\ \vdots \\ -f(\boldsymbol{e}_{:t+k})/(ka^-)
    \end{pmatrix}  \leq \mathbf{1}r \quad \forall (a^+, a^-) \in \mathcal{P}_j\label{eq:uaflexenvmincon} \\
    & \quad \ubar{p}-p^{\text{b}}_{l+t} \leq   r , \quad l=0,\dots, k-1.\nonumber
    \end{align}
    The left-hand side of \eqref{eq:uaflexenvmincon} is maximized for $a^-_{\max, j}$. We can then determine the value of $r$, and therefore $\ubar{\boldsymbol{R}}^{\alpha, k}_t$, as the maximum over all constraints, so
    \begin{equation*}
        \ubar{\boldsymbol{R}}^{\alpha, k}_t = \max \left\{\max_{l=1,\dots, k} \frac{-f(\boldsymbol{e}_{:t+l})}{a^-_{\max, j} l} , \max_{l=0,\dots, k-1}\ubar{p}-p^{\text{b}}_{t+l}\right\}
    \end{equation*}
    which proves the claim.

 \bibliographystyle{elsarticle-num} 
 \bibliography{references}





% \end{thebibliography}
\end{document}

%% file: flex_env.tex
\definecolor{crimson2143940}{RGB}{214,39,40}
\definecolor{darkgray176}{RGB}{176,176,176}
\definecolor{forestgreen4416044}{RGB}{44,160,44}
\definecolor{lightgray204}{RGB}{204,204,204}
\definecolor{steelblue31119180}{RGB}{31,119,180}
\begin{tikzpicture}[remember picture]

\begin{axis}[
width=250,
height=170,
legend cell align={left},
legend style={
  fill opacity=0.8,
  draw opacity=1,
  text opacity=1,
  at={(0.02,0.4)},
  anchor=west,
  draw=lightgray204
},
tick align=outside,
tick pos=left,
x grid style={darkgray176},
xlabel={Time [h]},
xmajorgrids,
xmin=-4.75, xmax=99.75,
xtick style={color=black},
xtick={0,24,48,72,96},
xticklabels={0,6,12,18,24},
y grid style={darkgray176},
ylabel={Power [kW]},
ymajorgrids,
ymin=-1.81601762504522, ymax=3.29210503093331,
ytick style={color=black}
]
\small
\addplot [line width=1.5pt, crimson2143940, opacity=0.7, dashed]
table {%
0 -1.48677232629561
1 -1.49334836478465
2 -1.49334008981773
3 -1.49333466292188
4 -1.49333110385042
5 -1.49332876973715
6 -1.493327238977
7 -1.49332623507258
8 -1.49332557669112
9 -1.49332514491084
10 -1.49332486174032
11 -1.4933246760312
12 -1.49332455423928
13 -1.4933244743656
14 -1.49332442198277
15 -1.49332438762902
16 -1.49332436509912
17 -1.49332435032354
18 -1.4933243406334
19 -1.49332433427841
20 -1.49332433011068
21 -1.49502611670374
22 -1.49672662935665
23 -1.49842486716618
24 -1.5001217000892
25 -1.53432197528748
26 -1.58383023159165
27 -1.48385410902999
28 -1.38506928197108
29 -1.27517500398318
30 -1.17789783827816
31 -1.10172562347304
32 -1.0354743252427
33 -1.01176034789573
34 -0.990323745031424
35 -0.971076887863804
36 -0.953905657783048
37 -0.895538055068707
38 -0.848346221878964
39 -0.809218011981818
40 -0.782100653571313
41 -0.776341723369459
42 -0.776341723369459
43 -0.776341723369459
44 -0.776341723369459
45 -0.776341723369459
46 -0.776341723369459
47 -0.776296810613437
48 -0.776296810613437
49 -0.776296810613437
50 -0.776296810613437
51 -0.776296810613437
52 -0.776296810613437
53 -0.776296810613437
54 -0.776296810613437
55 -0.776296810613437
56 -0.776296810613437
57 -0.776296810613437
58 -0.776296810613437
59 -0.778448532691298
60 -0.786324300194189
61 -0.866032932080771
62 -0.959607961255275
63 -1.0626750542674
64 -1.1596670501385
65 -1.20153680946243
66 -1.23162688945941
67 -1.25962803803907
68 -1.31339397825061
69 -1.34034460737316
70 -1.34034460737316
71 -1.34034460737316
72 -1.34034460737316
73 -1.34034460737316
74 -1.34034460737316
75 -1.34034460737316
76 -1.34034460737316
77 -1.34034460737316
78 -1.34034460737316
79 -1.34034460737316
80 -1.34034460737316
81 -1.32815524274247
82 -1.31005238622086
83 -1.29217177538599
84 -1.27451815756566
85 -1.26868477762699
86 -1.26868477762699
87 -1.26868477762699
88 -1.26868477762699
89 -1.28037721216326
90 -1.29810697702409
91 -1.31606223322232
92 -1.3342379901804
93 -1.34034460737316
94 -1.34034460737316
95 -1.34034460737316
};
\addlegendentry{Decrease potential}
\addplot [line width=1.5pt, forestgreen4416044, opacity=0.7, dashed]
table {%
0 2.02909117545057
1 2.02909117545057
2 2.02909117545057
3 2.02909117545057
4 2.02909117545057
5 2.06302647181266
6 2.09720658721377
7 2.13159374036143
8 2.16615037502265
9 2.20083922891046
10 2.23562340118151
11 2.27046641837187
12 2.30533229860556
13 2.30533229860556
14 2.30533229860556
15 2.30533229860556
16 2.30533229860556
17 2.30533229860556
18 2.30533229860556
19 2.30533229860556
20 2.30533229860556
21 2.30533229860556
22 2.30533229860556
23 2.30533229860556
24 2.30533229860556
25 2.30533229860556
26 2.30533229860556
27 2.30533229860556
28 2.30533229860556
29 2.30533229860556
30 2.30533229860556
31 2.30533229860556
32 2.30775286928531
33 2.31502687801247
34 2.32228400533532
35 2.32952876112407
36 2.34463871100455
37 2.4442035718419
38 2.54717147991408
39 2.64785663193418
40 2.73743106527374
41 2.7258842777914
42 2.73344220196503
43 2.74349441415308
44 2.75652401702
45 2.75113629311241
46 2.73332724046963
47 2.71931963227915
48 2.71361866190626
49 2.80823092027394
50 2.93605614415459
51 3.05991763747974
52 2.96425270620514
53 2.90879364195491
54 2.87668567094177
55 2.84953604125405
56 2.80936630128764
57 2.69695588472304
58 2.58382675370337
59 2.4731409400232
60 2.34403835364834
61 2.2849982839403
62 2.24918063028049
63 2.21340226333813
64 2.17770046563452
65 2.14309563411816
66 2.10864782732892
67 2.07439451110679
68 2.04037339945703
69 2.02909117545057
70 2.02909117545057
71 2.02909117545057
72 2.02909117545057
73 2.02909117545057
74 2.02909117545057
75 2.02909117545057
76 2.02909117545057
77 2.02909117545057
78 2.02909117545057
79 2.02909117545057
80 2.02909117545057
81 2.00662238483589
82 1.97317946734684
83 1.94008268303347
84 1.90737003145813
85 1.89655773056815
86 1.89655773056815
87 1.89655773056815
88 1.89655773056815
89 1.89655773056815
90 1.89655773056815
91 1.89655773056815
92 1.89655773056815
93 1.89655773056815
94 1.89655773056815
95 1.89655773056815
};
\addlegendentry{Increase potential}
\addplot [line width=1.5pt, steelblue31119180, opacity=0.7]
table {%
0 1.79311783938893
1 1.86282673681298
2 1.93229976297184
3 2.00124355078072
4 2.06935896254392
5 2.03542366618182
6 2.00124355078072
7 1.96685639763305
8 1.93229976297184
9 1.89761090908402
10 1.86282673681298
11 1.82798371962262
12 1.79311783938893
13 1.79311783938893
14 1.79311783938893
15 1.79311783938893
16 1.79311783938893
17 1.79311783938893
18 1.79311783938893
19 1.79311783938893
20 1.79311783938893
21 1.79311783938893
22 1.79311783938893
23 1.79311783938893
24 1.79311783938893
25 1.79311783938893
26 1.79311783938893
27 1.79311783938893
28 1.79311783938893
29 1.79311783938893
30 1.79311783938893
31 1.79311783938893
32 1.79069726870918
33 1.78342325998202
34 1.77616613265916
35 1.76892137687042
36 1.75381142698994
37 1.65424656615259
38 1.5512786580804
39 1.4505935060603
40 1.33363053996871
41 1.2467091902962
42 1.1514518122104
43 1.07848184825394
44 1.02731245809144
45 1.00436549709325
46 0.983669962567377
47 0.965128813485128
48 0.940713130623999
49 0.878522212182165
50 0.835060787252515
51 0.799417023845536
52 0.776341723369459
53 0.778279681148827
54 0.781037173366516
55 0.784909443391898
56 0.791881350900953
57 0.783870850887362
58 0.778433098188319
59 0.778850454830765
60 0.786324300194189
61 0.866032932080771
62 0.959607961255275
63 1.0626750542674
64 1.1596670501385
65 1.20153680946243
66 1.23162688945941
67 1.25962803803907
68 1.31339397825061
69 1.43954467888834
70 1.54952121038869
71 1.66621741960138
72 1.78956976487299
73 1.82539315690176
74 1.86120182601793
75 1.89695883675718
76 1.93229976297184
77 1.96685639763305
78 2.00124355078072
79 2.03542366618182
80 2.06935896254392
81 2.06935896254392
82 2.06935896254392
83 2.06935896254392
84 2.06935896254392
85 2.06935896254392
86 2.06935896254392
87 2.06935896254392
88 2.06935896254392
89 2.06935896254392
90 2.06935896254392
91 2.06935896254392
92 2.06935896254392
93 2.10301150361272
94 2.13634326929799
95 2.16931622764019
};
\addlegendentry{Baseline}
\end{axis}

\end{tikzpicture}

%% file: self_con_3days.tex
% This file was created with tikzplotlib v0.10.1.
\begin{tikzpicture}

\definecolor{crimson2143940}{RGB}{214,39,40}
\definecolor{darkgray176}{RGB}{176,176,176}
\definecolor{darkorange25512714}{RGB}{255,127,14}
\definecolor{forestgreen4416044}{RGB}{44,160,44}
\definecolor{lightgray204}{RGB}{204,204,204}
\definecolor{steelblue31119180}{RGB}{31,119,180}

\begin{axis}[
width=250,
height=150,
legend cell align={left},
legend style={fill opacity=0.8, draw opacity=1, text opacity=1, draw=lightgray204},
tick align=outside,
tick pos=left,
x grid style={darkgray176},
xlabel={Time [h]},
xmajorgrids,
xmin=-3.58824755157324, xmax=75.3531985830381,
xtick style={color=black},
y grid style={darkgray176},
ylabel={Power [kW]},
ymajorgrids,
ymin=43.3153331662146, ymax=345.034285540106,
ytick style={color=black}
]
\small
\addplot [line width=1.5pt, steelblue31119180, opacity=0.7]
table {%
0 142.622944032442
0.250052094186289 148.167100215968
0.500104188372578 153.693115765988
0.750156282558866 159.177661705829
1.00020837674516 164.596917476617
1.25026047093144 161.89695414767
1.50031256511773 159.177661705829
1.75036465930402 156.44205170039
2.00041675349031 153.693115765988
2.2504688476766 150.933820126414
2.50052094186289 148.167100215968
2.75057303604918 145.395855432082
3.00062513023547 142.622944032442
3.25067722442175 142.622944032442
3.50072931860804 142.622944032442
3.75078141279433 142.622944032442
4.00083350698062 142.622944032442
4.25088560116691 142.622944032442
4.5009376953532 142.622944032442
4.75098978953949 142.622944032442
5.00104188372578 142.622944032442
5.25109397791207 142.622944032442
5.50114607209835 142.622944032442
5.75119816628464 142.622944032442
6.00125026047093 142.622944032442
6.25130235465722 142.622944032442
6.50135444884351 142.622944032442
6.7514065430298 142.622944032442
7.00145863721609 142.622944032442
7.25151073140238 142.622944032442
7.50156282558866 142.622944032442
7.75161491977495 142.622944032442
8.00166701396124 142.406815936456
8.25171910814753 141.757442362922
8.50177120233382 141.109738883884
8.75182329652011 140.463302747331
9.0018753907064 139.115598077739
9.25192748489269 130.911129284556
9.50197957907898 122.404192793551
9.75203167326526 114.09844530895
10.0020837674516 104.343042581695
10.2521358616378 96.6413031069522
10.5021879558241 126.276103594776
10.7522400500104 156.919735827252
11.0022921441967 175.572948101988
11.252344238383 187.921696600677
11.5023963325693 199.990009636223
11.7524484267556 211.747398182144
12.0025005209419 237.111634912603
12.2525526151282 269.636304857455
12.5026047093144 301.20489352539
12.7526568035007 328.421549724643
13.002708897687 331.319787704929
13.2527609918733 322.898264033306
13.5028130860596 313.042206097803
13.7528651802459 301.746700042967
14.0029172744322 282.699422368649
14.2529693686185 260.276537566648
14.5030214628048 238.354224508914
14.753073556991 214.174608367824
15.0031256511773 182.357155786813
15.2531777453636 146.85608718593
15.5032298395499 114.033683570633
15.7532819337362 84.13920820247
16.0033340279225 90.1316581829145
16.2533861221088 93.8461036800206
16.5034382162951 96.5184862696345
16.7534903104814 99.0073629499101
17.0035424046676 103.791311302274
17.2535944988539 114.080831786111
17.5036465930402 122.905492563277
17.7536986872265 132.312674617655
18.0037507814128 142.306149446677
18.2538028755991 145.164535579421
18.5038549697854 148.021996060354
18.7539070639717 150.87558628216
19.003959158158 153.693115765988
19.2540112523442 156.44205170039
19.5040633465305 159.177661705829
19.7541154407168 161.89695414767
20.0041675349031 164.596917476617
20.2542196290894 164.596917476617
20.5042717232757 164.596917476617
20.754323817462 164.596917476617
21.0043759116483 164.596917476617
21.2544280058346 164.596917476617
21.5044801000208 164.596917476617
21.7545321942071 164.596917476617
22.0045842883934 164.596917476617
22.2546363825797 164.596917476617
22.504688476766 164.596917476617
22.7547405709523 164.596917476617
23.0047926651386 164.596917476617
23.2548447593249 167.274525828186
23.5048968535111 169.926744710828
23.7549489476974 172.550536767675
24.0050010418837 175.142867596536
24.25505313607 175.142867596536
24.5051052302563 175.142867596536
24.7551573244426 175.142867596536
25.0052094186289 175.142867596536
25.2552615128152 172.550536767675
25.5053136070015 169.926744710828
25.7553657011877 167.274525828186
26.005417795374 164.596917476617
26.2554698895603 164.596917476617
26.5055219837466 164.596917476617
26.7555740779329 164.596917476617
27.0056261721192 164.596917476617
27.2556782663055 164.596917476617
27.5057303604918 164.596917476617
27.7557824546781 164.596917476617
28.0058345488643 164.596917476617
28.2558866430506 164.596917476617
28.5059387372369 164.596917476617
28.7559908314232 164.596917476617
29.0060429256095 164.596917476617
29.2560950197958 164.596917476617
29.5061471139821 164.596917476617
29.7561992081684 164.596917476617
30.0062513023547 164.596917476617
30.2563033965409 164.596917476617
30.5063554907272 164.596917476617
30.7564075849135 164.596917476617
31.0064596790998 164.596917476617
31.2565117732861 161.89695414767
31.5065638674724 159.177661705829
31.7566159616587 156.44205170039
32.006668055845 153.414819313231
32.2567201500313 152.578235405257
32.5067722442175 151.743892299586
32.7568243384038 150.911247961444
33.0068764325901 150.269939638295
33.2569285267764 147.263429098475
33.5069806209627 144.252715743814
33.757032715149 141.241109068902
34.0070848093353 137.74742258059
34.2571369035216 127.857112436387
34.5071889977079 118.147466574183
34.7572410918941 108.702604094384
35.0072931860804 99.3396424742371
35.2573452802667 96.9879975088368
35.507397374453 94.6703883251952
35.7574494686393 92.3846213404893
36.0075015628256 91.2948184494642
36.2575536570119 92.5084279809108
36.5076057511982 93.6913104811587
36.7576578453845 94.8524621530309
37.0077099395707 94.1809459496704
37.257762033757 87.5641100494375
37.5078141279433 92.4223117629385
37.7578662221296 109.809747821702
38.0079183163159 114.820397111927
38.2579704105022 113.609797954169
38.5080225046885 112.53683150534
38.7580745988748 111.613670085003
39.0081266930611 98.580172533267
39.2581787872473 79.9106932067403
39.5082308814336 81.2951018951579
39.7582829756199 85.8478714016391
40.0083350698062 89.246207452422
40.2583871639925 90.1342014106676
40.5084392581788 91.0366995140715
40.7584913523651 92.3652653361395
41.0085434465514 94.8483709263876
41.2585955407377 99.5693908630134
41.5086476349239 102.471294743169
41.7586997291102 106.047648397727
42.0087518232965 109.74315194448
42.2588039174828 109.863039893803
42.5088560116691 109.982805354801
42.7589081058554 110.102445636579
43.0089602000417 110.18213510652
43.259012294228 112.773266780943
43.5090643884143 115.39221190858
43.7591164826005 118.036936034875
44.0091685767868 120.705317539315
44.2592206709731 118.036936034875
44.5092727651594 115.39221190858
44.7593248593457 112.773266780943
45.009376953532 110.18213510652
45.2594290477183 112.773266780943
45.5094811419046 115.39221190858
45.7595332360909 118.036936034875
46.0095853302771 120.705317539315
46.2596374244634 120.705317539315
46.5096895186497 120.705317539315
46.759741612836 120.705317539315
47.0097937070223 120.705317539315
47.2598458012086 120.705317539315
47.5098978953949 120.705317539315
47.7599499895812 120.705317539315
48.0100020837675 120.705317539315
48.2600541779537 123.395150212826
48.51010627214 126.104146084836
48.7601583663263 128.829938495581
49.0102104605126 131.570085408386
49.2602625546989 131.570085408386
49.5103146488852 131.570085408386
49.7603667430715 131.570085408386
50.0104188372578 131.570085408386
50.2604709314441 131.570085408386
50.5105230256303 131.570085408386
50.7605751198166 131.570085408386
51.0106272140029 131.570085408386
51.2606793081892 134.322072955838
51.5107314023755 137.083319212904
51.7607834965618 139.851178189282
52.0108355907481 142.622944032442
52.2608876849344 148.167100215968
52.5109397791207 153.693115765988
52.7609918733069 159.177661705829
53.0110439674932 164.596917476617
53.2610960616795 161.89695414767
53.5111481558658 159.177661705829
53.7612002500521 156.44205170039
54.0112523442384 153.693115765988
54.2613044384247 156.44205170039
54.511356532611 159.177661705829
54.7614086267972 161.89695414767
55.0114607209835 164.596917476617
55.2615128151698 164.596917476617
55.5115649093561 164.596917476617
55.7616170035424 164.596917476617
56.0116690977287 164.374816609588
56.261721191915 163.707257017644
56.5117732861013 163.040584666967
56.7618253802876 162.37451010648
57.0118774744738 161.147212092677
57.2619295686601 158.772546012982
57.5119816628464 156.382877657021
57.7620337570327 153.837357440097
58.012085851219 151.301680575628
58.2621379454053 143.698679218808
58.5121900395916 136.142188668557
58.7622421337779 128.664810273689
59.0122942279642 121.067346769317
59.2623463221504 115.751946079852
59.5123984163367 110.55426439282
59.762450510523 105.501749579148
60.0125026047093 110.643978551412
60.2625546988956 124.093754018967
60.5126067930819 137.60331789604
60.7626588872682 151.175929507391
61.0127109814545 154.564147645366
61.2627630756408 152.850570074908
61.512815169827 151.123175228954
61.7628672640133 149.357746658842
62.0129193581996 139.46322111003
62.2629714523859 126.071595042048
62.5130235465722 113.52552039977
62.7630756407585 101.880212365814
63.0131277349448 94.1288578908444
63.2631798291311 96.9371337799784
63.5132319233174 101.063709905655
63.7632840175037 105.365182490325
64.0133361116899 110.203862135075
64.2633882058762 113.241228703384
64.5134403000625 116.286619528259
64.7634923942488 120.007551305728
65.0135444884351 124.711505014977
65.2635965826214 136.191098920524
65.5136486768077 144.955673296906
65.763700770994 154.442284892037
66.0137528651803 164.108297276908
66.2638049593665 164.241825300985
66.5138570535528 164.375153794946
66.7639091477391 164.508280089738
67.0139612419254 164.596917476617
67.2640133361117 169.926744710828
67.514065430298 175.142867596536
67.7641175244843 180.221057937447
68.0141696186706 185.137322930408
68.2642217128569 185.137322930408
68.5142738070431 185.137322930408
68.7643259012294 185.137322930408
69.0143779954157 185.137322930408
69.264430089602 185.137322930408
69.5144821837883 185.137322930408
69.7645342779746 185.137322930408
70.0145863721609 185.137322930408
70.2646384663472 185.137322930408
70.5146905605335 185.137322930408
70.7647426547197 185.137322930408
71.014794748906 185.137322930408
71.2648468430923 185.137322930408
71.5148989372786 185.137322930408
71.7649510314649 185.137322930408
};
\addlegendentry{baseline + request}
\addplot [line width=1.5pt, crimson2143940, opacity=0.7, dash pattern=on 1pt off 3pt on 3pt off 3pt]
table {%
0 142.622944032442
0.250052094186289 148.167100215968
0.500104188372578 153.693115765988
0.750156282558866 159.177661705829
1.00020837674516 164.596917476617
1.25026047093144 161.89695414767
1.50031256511773 159.177661705829
1.75036465930402 156.44205170039
2.00041675349031 153.693115765988
2.2504688476766 150.933820126414
2.50052094186289 148.167100215968
2.75057303604918 145.395855432082
3.00062513023547 142.622944032442
3.25067722442175 142.622944032442
3.50072931860804 142.622944032442
3.75078141279433 142.622944032442
4.00083350698062 142.622944032442
4.25088560116691 142.622944032442
4.5009376953532 142.622944032442
4.75098978953949 142.622944032442
5.00104188372578 142.622944032442
5.25109397791207 142.622944032442
5.50114607209835 142.622944032442
5.75119816628464 142.622944032442
6.00125026047093 142.622944032442
6.25130235465722 142.622944032442
6.50135444884351 142.622944032442
6.7514065430298 142.622944032442
7.00145863721609 142.622944032442
7.25151073140238 142.622944032442
7.50156282558866 142.622944032442
7.75161491977495 142.622944032442
8.00166701396124 142.406815936456
8.25171910814753 141.757442362922
8.50177120233382 141.109738883884
8.75182329652011 140.463302747331
9.0018753907064 139.115598077739
9.25192748489269 130.911129284556
9.50197957907898 122.404192793551
9.75203167326526 114.09844530895
10.0020837674516 104.343042581695
10.2521358616378 96.6413031069522
10.5021879558241 88.2367179368538
10.7522400500104 81.835195650429
11.0022921441967 138.986668872733
11.252344238383 140.403953283925
11.5023963325693 138.618134519912
11.7524484267556 137.026289741506
12.0025005209419 141.502782221076
12.2525526151282 178.857562795514
12.5026047093144 215.969777554218
12.7526568035007 251.705693938524
13.002708897687 252.66992940615
13.2527609918733 252.792534245838
13.5028130860596 252.980404676723
13.7528651802459 253.260475379572
14.0029172744322 192.180531328371
14.2529693686185 188.207986964503
14.5030214628048 187.892278606505
14.753073556991 188.106340313071
15.0031256511773 182.357155786813
15.2531777453636 146.713367921849
15.5032298395499 114.033683570633
15.7532819337362 84.13920820247
16.0033340279225 90.1316581829145
16.2533861221088 93.8461036800206
16.5034382162951 96.5184862696345
16.7534903104814 99.0073629499101
17.0035424046676 103.791311302274
17.2535944988539 114.080831786111
17.5036465930402 122.905492563277
17.7536986872265 132.312674617655
18.0037507814128 142.306149446677
18.2538028755991 145.164535579421
18.5038549697854 148.021996060354
18.7539070639717 150.87558628216
19.003959158158 153.693115765988
19.2540112523442 156.44205170039
19.5040633465305 159.177661705829
19.7541154407168 161.89695414767
20.0041675349031 164.596917476617
20.2542196290894 164.596917476617
20.5042717232757 164.596917476617
20.754323817462 164.596917476617
21.0043759116483 164.596917476617
21.2544280058346 164.596917476617
21.5044801000208 164.596917476617
21.7545321942071 164.596917476617
22.0045842883934 164.596917476617
22.2546363825797 164.596917476617
22.504688476766 164.596917476617
22.7547405709523 164.596917476617
23.0047926651386 164.596917476617
23.2548447593249 167.274525828186
23.5048968535111 169.926744710828
23.7549489476974 172.550536767675
24.0050010418837 175.142867596536
24.25505313607 175.142867596536
24.5051052302563 175.142867596536
24.7551573244426 175.142867596536
25.0052094186289 175.142867596536
25.2552615128152 172.550536767675
25.5053136070015 169.926744710828
25.7553657011877 167.274525828186
26.005417795374 164.596917476617
26.2554698895603 164.596917476617
26.5055219837466 164.596917476617
26.7555740779329 164.596917476617
27.0056261721192 164.596917476617
27.2556782663055 164.596917476617
27.5057303604918 164.596917476617
27.7557824546781 164.596917476617
28.0058345488643 164.596917476617
28.2558866430506 164.596917476617
28.5059387372369 164.596917476617
28.7559908314232 164.596917476617
29.0060429256095 164.596917476617
29.2560950197958 164.596917476617
29.5061471139821 164.596917476617
29.7561992081684 164.596917476617
30.0062513023547 164.596917476617
30.2563033965409 164.596917476617
30.5063554907272 164.596917476617
30.7564075849135 164.596917476617
31.0064596790998 164.596917476617
31.2565117732861 161.89695414767
31.5065638674724 159.177661705829
31.7566159616587 156.44205170039
32.006668055845 153.414819313231
32.2567201500313 152.578235405257
32.5067722442175 151.743892299586
32.7568243384038 150.911247961444
33.0068764325901 150.269939638295
33.2569285267764 147.263429098475
33.5069806209627 144.252715743814
33.757032715149 141.241109068902
34.0070848093353 137.74742258059
34.2571369035216 127.857112436387
34.5071889977079 118.147466574183
34.7572410918941 108.702604094384
35.0072931860804 99.3396424742371
35.2573452802667 96.9879975088368
35.507397374453 94.6703883251952
35.7574494686393 92.3846213404893
36.0075015628256 91.2948184494642
36.2575536570119 92.5084279809108
36.5076057511982 93.6913104811587
36.7576578453845 94.8524621530309
37.0077099395707 94.1809459496704
37.257762033757 87.5641100494375
37.5078141279433 92.4223117629385
37.7578662221296 109.809747821702
38.0079183163159 114.820397111927
38.2579704105022 113.609797954169
38.5080225046885 112.53683150534
38.7580745988748 111.613670085003
39.0081266930611 98.580172533267
39.2581787872473 79.9106932067403
39.5082308814336 81.2951018951579
39.7582829756199 85.8478714016391
40.0083350698062 89.246207452422
40.2583871639925 90.1342014106676
40.5084392581788 91.0366995140715
40.7584913523651 92.3652653361395
41.0085434465514 94.8483709263876
41.2585955407377 99.5693908630134
41.5086476349239 102.471294743169
41.7586997291102 106.047648397727
42.0087518232965 109.74315194448
42.2588039174828 109.863039893803
42.5088560116691 109.982805354801
42.7589081058554 110.102445636579
43.0089602000417 110.18213510652
43.259012294228 112.773266780943
43.5090643884143 115.39221190858
43.7591164826005 118.036936034875
44.0091685767868 120.705317539315
44.2592206709731 118.036936034875
44.5092727651594 115.39221190858
44.7593248593457 112.773266780943
45.009376953532 110.18213510652
45.2594290477183 112.773266780943
45.5094811419046 115.39221190858
45.7595332360909 118.036936034875
46.0095853302771 120.705317539315
46.2596374244634 120.705317539315
46.5096895186497 120.705317539315
46.759741612836 120.705317539315
47.0097937070223 120.705317539315
47.2598458012086 120.705317539315
47.5098978953949 120.705317539315
47.7599499895812 120.705317539315
48.0100020837675 120.705317539315
48.2600541779537 123.395150212826
48.51010627214 126.104146084836
48.7601583663263 128.829938495581
49.0102104605126 131.570085408386
49.2602625546989 131.570085408386
49.5103146488852 131.570085408386
49.7603667430715 131.570085408386
50.0104188372578 131.570085408386
50.2604709314441 131.570085408386
50.5105230256303 131.570085408386
50.7605751198166 131.570085408386
51.0106272140029 131.570085408386
51.2606793081892 134.322072955838
51.5107314023755 137.083319212904
51.7607834965618 139.851178189282
52.0108355907481 142.622944032442
52.2608876849344 148.167100215968
52.5109397791207 153.693115765988
52.7609918733069 159.177661705829
53.0110439674932 164.596917476617
53.2610960616795 161.89695414767
53.5111481558658 159.177661705829
53.7612002500521 156.44205170039
54.0112523442384 153.693115765988
54.2613044384247 156.44205170039
54.511356532611 159.177661705829
54.7614086267972 161.89695414767
55.0114607209835 164.596917476617
55.2615128151698 164.596917476617
55.5115649093561 164.596917476617
55.7616170035424 164.596917476617
56.0116690977287 164.374816609588
56.261721191915 163.707257017644
56.5117732861013 163.040584666967
56.7618253802876 162.37451010648
57.0118774744738 161.147212092677
57.2619295686601 158.772546012982
57.5119816628464 156.382877657021
57.7620337570327 153.837357440097
58.012085851219 151.301680575628
58.2621379454053 143.698679218808
58.5121900395916 136.142188668557
58.7622421337779 128.664810273689
59.0122942279642 121.067346769317
59.2623463221504 115.751946079852
59.5123984163367 110.55426439282
59.762450510523 105.501749579148
60.0125026047093 110.643978551412
60.2625546988956 124.093754018967
60.5126067930819 137.60331789604
60.7626588872682 151.175929507391
61.0127109814545 154.564147645366
61.2627630756408 152.850570074908
61.512815169827 151.123175228954
61.7628672640133 149.357746658842
62.0129193581996 139.46322111003
62.2629714523859 126.071595042048
62.5130235465722 113.52552039977
62.7630756407585 101.880212365814
63.0131277349448 94.1288578908444
63.2631798291311 96.9371337799784
63.5132319233174 101.063709905655
63.7632840175037 105.365182490325
64.0133361116899 110.203862135075
64.2633882058762 113.241228703384
64.5134403000625 116.286619528259
64.7634923942488 120.007551305728
65.0135444884351 124.711505014977
65.2635965826214 136.191098920524
65.5136486768077 144.955673296906
65.763700770994 154.442284892037
66.0137528651803 164.108297276908
66.2638049593665 164.241825300985
66.5138570535528 164.375153794946
66.7639091477391 164.508280089738
67.0139612419254 164.596917476617
67.2640133361117 169.926744710828
67.514065430298 175.142867596536
67.7641175244843 180.221057937447
68.0141696186706 185.137322930408
68.2642217128569 185.137322930408
68.5142738070431 185.137322930408
68.7643259012294 185.137322930408
69.0143779954157 185.137322930408
69.264430089602 185.137322930408
69.5144821837883 185.137322930408
69.7645342779746 185.137322930408
70.0145863721609 185.137322930408
70.2646384663472 185.137322930408
70.5146905605335 185.137322930408
70.7647426547197 185.137322930408
71.014794748906 185.137322930408
71.2648468430923 185.137322930408
71.5148989372786 185.137322930408
71.7649510314649 185.137322930408
};
\addlegendentry{committed trajectory}
\addplot [line width=1.5pt, forestgreen4416044, opacity=0.7, dashed]
table {%
0 142.622944032442
0.250052094186289 148.167100215968
0.500104188372578 153.693115765988
0.750156282558866 159.177661705829
1.00020837674516 164.596917476617
1.25026047093144 161.89695414767
1.50031256511773 159.177661705829
1.75036465930402 156.44205170039
2.00041675349031 153.693115765988
2.2504688476766 150.933820126414
2.50052094186289 148.167100215968
2.75057303604918 145.395855432082
3.00062513023547 142.622944032442
3.25067722442175 142.622944032442
3.50072931860804 142.622944032442
3.75078141279433 142.622944032442
4.00083350698062 142.622944032442
4.25088560116691 142.622944032442
4.5009376953532 142.622944032442
4.75098978953949 142.622944032442
5.00104188372578 142.622944032442
5.25109397791207 142.622944032442
5.50114607209835 142.622944032442
5.75119816628464 142.622944032442
6.00125026047093 142.622944032442
6.25130235465722 142.622944032442
6.50135444884351 142.622944032442
6.7514065430298 142.622944032442
7.00145863721609 142.622944032442
7.25151073140238 142.622944032442
7.50156282558866 142.622944032442
7.75161491977495 142.622944032442
8.00166701396124 142.406815936456
8.25171910814753 141.757442362922
8.50177120233382 141.109738883884
8.75182329652011 140.463302747331
9.0018753907064 139.115598077739
9.25192748489269 130.911129284556
9.50197957907898 122.404192793551
9.75203167326526 114.09844530895
10.0020837674516 104.343042581695
10.2521358616378 96.6413031069522
10.5021879558241 88.2367179368538
10.7522400500104 81.835195650429
11.0022921441967 77.3751871440063
11.252344238383 75.3862743428356
11.5023963325693 73.60045669289
11.7524484267556 72.0086118792102
12.0025005209419 69.9277790909075
12.2525526151282 64.8768190207902
12.5026047093144 61.4539248645217
12.7526568035007 58.7463856421718
13.002708897687 57.0298310013915
13.2527609918733 57.1524339308298
13.5028130860596 57.3403059734313
13.7528651802459 57.6203757363227
14.0029172744322 58.1519138637251
14.2529693686185 57.5855657504985
14.5030214628048 57.2698569726047
14.753073556991 57.4839188790607
15.0031256511773 58.3396907238823
15.2531777453636 65.0540111261034
15.5032298395499 73.0150895623103
15.7532819337362 81.8349353111378
16.0033340279225 90.1316581829145
16.2533861221088 93.8461036800206
16.5034382162951 96.5184862696345
16.7534903104814 99.0073629499101
17.0035424046676 103.791311302274
17.2535944988539 114.080831786111
17.5036465930402 122.905492563277
17.7536986872265 132.312674617655
18.0037507814128 142.306149446677
18.2538028755991 145.164535579421
18.5038549697854 148.021996060354
18.7539070639717 150.87558628216
19.003959158158 153.693115765988
19.2540112523442 156.44205170039
19.5040633465305 159.177661705829
19.7541154407168 161.89695414767
20.0041675349031 164.596917476617
20.2542196290894 164.596917476617
20.5042717232757 164.596917476617
20.754323817462 164.596917476617
21.0043759116483 164.596917476617
21.2544280058346 164.596917476617
21.5044801000208 164.596917476617
21.7545321942071 164.596917476617
22.0045842883934 164.596917476617
22.2546363825797 164.596917476617
22.504688476766 164.596917476617
22.7547405709523 164.596917476617
23.0047926651386 164.596917476617
23.2548447593249 167.274525828186
23.5048968535111 169.926744710828
23.7549489476974 172.550536767675
24.0050010418837 175.142867596536
24.25505313607 175.142867596536
24.5051052302563 175.142867596536
24.7551573244426 175.142867596536
25.0052094186289 175.142867596536
25.2552615128152 172.550536767675
25.5053136070015 169.926744710828
25.7553657011877 167.274525828186
26.005417795374 164.596917476617
26.2554698895603 164.596917476617
26.5055219837466 164.596917476617
26.7555740779329 164.596917476617
27.0056261721192 164.596917476617
27.2556782663055 164.596917476617
27.5057303604918 164.596917476617
27.7557824546781 164.596917476617
28.0058345488643 164.596917476617
28.2558866430506 164.596917476617
28.5059387372369 164.596917476617
28.7559908314232 164.596917476617
29.0060429256095 164.596917476617
29.2560950197958 164.596917476617
29.5061471139821 164.596917476617
29.7561992081684 164.596917476617
30.0062513023547 164.596917476617
30.2563033965409 164.596917476617
30.5063554907272 164.596917476617
30.7564075849135 164.596917476617
31.0064596790998 164.596917476617
31.2565117732861 161.89695414767
31.5065638674724 159.177661705829
31.7566159616587 156.44205170039
32.006668055845 153.414819313231
32.2567201500313 152.578235405257
32.5067722442175 151.743892299586
32.7568243384038 150.911247961444
33.0068764325901 150.269939638295
33.2569285267764 147.263429098475
33.5069806209627 144.252715743814
33.757032715149 141.241109068902
34.0070848093353 137.74742258059
34.2571369035216 127.857112436387
34.5071889977079 118.147466574183
34.7572410918941 108.702604094384
35.0072931860804 99.3396424742371
35.2573452802667 96.9879975088368
35.507397374453 94.6703883251952
35.7574494686393 92.3846213404893
36.0075015628256 91.2948184494642
36.2575536570119 92.5084279809108
36.5076057511982 93.6913104811587
36.7576578453845 94.8524621530309
37.0077099395707 94.1809459496704
37.257762033757 87.5641100494375
37.5078141279433 81.1901558373563
37.7578662221296 75.111237517966
38.0079183163159 70.6467175452646
38.2579704105022 70.8972868758102
38.5080225046885 71.1223859149676
38.7580745988748 71.3195332916588
39.0081266930611 72.761439694605
39.2581787872473 76.9165899784554
39.5082308814336 81.2951018951579
39.7582829756199 85.8478714016391
40.0083350698062 89.246207452422
40.2583871639925 90.1342014106676
40.5084392581788 91.0366995140715
40.7584913523651 92.3652653361395
41.0085434465514 94.8483709263876
41.2585955407377 99.5693908630134
41.5086476349239 102.471294743169
41.7586997291102 106.047648397727
42.0087518232965 109.74315194448
42.2588039174828 109.863039893803
42.5088560116691 109.982805354801
42.7589081058554 110.102445636579
43.0089602000417 110.18213510652
43.259012294228 112.773266780943
43.5090643884143 115.39221190858
43.7591164826005 118.036936034875
44.0091685767868 120.705317539315
44.2592206709731 118.036936034875
44.5092727651594 115.39221190858
44.7593248593457 112.773266780943
45.009376953532 110.18213510652
45.2594290477183 112.773266780943
45.5094811419046 115.39221190858
45.7595332360909 118.036936034875
46.0095853302771 120.705317539315
46.2596374244634 120.705317539315
46.5096895186497 120.705317539315
46.759741612836 120.705317539315
47.0097937070223 120.705317539315
47.2598458012086 120.705317539315
47.5098978953949 120.705317539315
47.7599499895812 120.705317539315
48.0100020837675 120.705317539315
48.2600541779537 123.395150212826
48.51010627214 126.104146084836
48.7601583663263 128.829938495581
49.0102104605126 131.570085408386
49.2602625546989 131.570085408386
49.5103146488852 131.570085408386
49.7603667430715 131.570085408386
50.0104188372578 131.570085408386
50.2604709314441 131.570085408386
50.5105230256303 131.570085408386
50.7605751198166 131.570085408386
51.0106272140029 131.570085408386
51.2606793081892 134.322072955838
51.5107314023755 137.083319212904
51.7607834965618 139.851178189282
52.0108355907481 142.622944032442
52.2608876849344 148.167100215968
52.5109397791207 153.693115765988
52.7609918733069 159.177661705829
53.0110439674932 164.596917476617
53.2610960616795 161.89695414767
53.5111481558658 159.177661705829
53.7612002500521 156.44205170039
54.0112523442384 153.693115765988
54.2613044384247 156.44205170039
54.511356532611 159.177661705829
54.7614086267972 161.89695414767
55.0114607209835 164.596917476617
55.2615128151698 164.596917476617
55.5115649093561 164.596917476617
55.7616170035424 164.596917476617
56.0116690977287 164.374816609588
56.261721191915 163.707257017644
56.5117732861013 163.040584666967
56.7618253802876 162.37451010648
57.0118774744738 161.147212092677
57.2619295686601 158.772546012982
57.5119816628464 156.382877657021
57.7620337570327 153.837357440097
58.012085851219 151.301680575628
58.2621379454053 143.698679218808
58.5121900395916 136.142188668557
58.7622421337779 128.664810273689
59.0122942279642 121.067346769317
59.2623463221504 115.751946079852
59.5123984163367 110.55426439282
59.762450510523 105.501749579148
60.0125026047093 100.277007540911
60.2625546988956 97.0083636493862
60.5126067930819 93.9093174402447
60.7626588872682 90.9867304289091
61.0127109814545 89.2530465351144
61.2627630756408 89.5968006931407
61.512815169827 89.944511726924
61.7628672640133 90.3017437823125
62.0129193581996 91.5251062920875
62.2629714523859 92.127299016027
62.5130235465722 92.6555406499045
62.7630756407585 93.0748382370399
63.0131277349448 92.9721186414502
63.2631798291311 96.9371337799784
63.5132319233174 101.063709905655
63.7632840175037 105.365182490325
64.0133361116899 110.203862135075
64.2633882058762 113.241228703384
64.5134403000625 116.286619528259
64.7634923942488 120.007551305728
65.0135444884351 124.711505014977
65.2635965826214 136.191098920524
65.5136486768077 144.955673296906
65.763700770994 154.442284892037
66.0137528651803 164.108297276908
66.2638049593665 164.241825300985
66.5138570535528 164.375153794946
66.7639091477391 164.508280089738
67.0139612419254 164.596917476617
67.2640133361117 169.926744710828
67.514065430298 175.142867596536
67.7641175244843 180.221057937447
68.0141696186706 185.137322930408
68.2642217128569 185.137322930408
68.5142738070431 185.137322930408
68.7643259012294 185.137322930408
69.0143779954157 185.137322930408
69.264430089602 185.137322930408
69.5144821837883 185.137322930408
69.7645342779746 185.137322930408
70.0145863721609 185.137322930408
70.2646384663472 185.137322930408
70.5146905605335 185.137322930408
70.7647426547197 185.137322930408
71.014794748906 185.137322930408
71.2648468430923 185.137322930408
71.5148989372786 185.137322930408
71.7649510314649 185.137322930408
};
\addlegendentry{baseline consumption}
\addplot [line width=1.5pt, darkorange25512714, opacity=0.7, dotted]
table {%
0 143.505340893826
0.250052094186289 146.48055968217
0.500104188372578 159.078129152104
0.750156282558866 169.015243829517
1.00020837674516 173.371169910756
1.25026047093144 170.586534495652
1.50031256511773 162.122798048013
1.75036465930402 157.188311089956
2.00041675349031 152.468253078685
2.2504688476766 148.60681760641
2.50052094186289 144.716605185527
2.75057303604918 142.63500356814
3.00062513023547 140.676473797556
3.25067722442175 140.065941051306
3.50072931860804 142.160428728968
3.75078141279433 143.231780983627
4.00083350698062 145.376768136304
4.25088560116691 146.311761868128
4.5009376953532 145.370806202206
4.75098978953949 144.260006661485
5.00104188372578 144.017468345341
5.25109397791207 144.568614448435
5.50114607209835 145.092552191437
5.75119816628464 144.890737706934
6.00125026047093 144.247127088576
6.25130235465722 144.048869230923
6.50135444884351 144.184234879933
6.7514065430298 144.963106592681
7.00145863721609 144.944440600944
7.25151073140238 144.42997074947
7.50156282558866 144.502456425197
7.75161491977495 144.475743502755
8.00166701396124 144.519901371752
8.25171910814753 118.598825323312
8.50177120233382 117.94410552571
8.75182329652011 118.039807154942
9.0018753907064 120.667664613299
9.25192748489269 118.769533264977
9.50197957907898 108.466728150357
9.75203167326526 99.449275553491
10.0020837674516 92.3499158757205
10.2521358616378 89.2763059193607
10.5021879558241 88.2367179368538
10.7522400500104 81.835195650429
11.0022921441967 138.986668872733
11.252344238383 140.403953283925
11.5023963325693 138.618134519912
11.7524484267556 137.026289741506
12.0025005209419 141.502782221076
12.2525526151282 178.857562795514
12.5026047093144 215.969777554218
12.7526568035007 251.705693938524
13.002708897687 252.66992940615
13.2527609918733 252.792534245838
13.5028130860596 252.980404676723
13.7528651802459 253.260475379572
14.0029172744322 192.180531328371
14.2529693686185 188.207986964503
14.5030214628048 187.892278606505
14.753073556991 188.106340313071
15.0031256511773 182.357155786814
15.2531777453636 146.713367921849
15.5032298395499 114.033683570633
15.7532819337362 84.13920820247
16.0033340279225 72.1053265463316
16.2533861221088 75.0768829440165
16.5034382162951 77.2147890157076
16.7534903104814 79.205890359928
17.0035424046676 83.0330490418195
17.2535944988539 91.2646654288889
17.5036465930402 98.3243940506216
17.7536986872265 105.850139694124
18.0037507814128 113.844919557342
18.2538028755991 116.131628463537
18.5038549697854 118.417596848283
18.7539070639717 121.291958528527
19.003959158158 124.861489731146
19.2540112523442 127.094104591405
19.5040633465305 130.258220217563
19.7541154407168 131.818479622558
20.0041675349031 134.851125995083
20.2542196290894 133.811275605351
20.5042717232757 136.706230494492
20.754323817462 137.505611889092
21.0043759116483 139.743392513007
21.2544280058346 140.906554254267
21.5044801000208 138.228979806705
21.7545321942071 140.404968218304
22.0045842883934 140.280136950083
22.2546363825797 142.425954369138
22.504688476766 142.997668123682
22.7547405709523 141.115263295543
23.0047926651386 141.696039896309
23.2548447593249 145.438946902798
23.5048968535111 151.093631015285
23.7549489476974 154.11079785676
24.0050010418837 156.671912753956
24.25505313607 159.118013565275
24.5051052302563 160.994836532593
24.7551573244426 160.723044856187
25.0052094186289 159.072329553538
25.2552615128152 155.339747645182
25.5053136070015 153.133659524918
25.7553657011877 150.497914609098
26.005417795374 148.531586951234
26.2554698895603 149.263125284592
26.5055219837466 151.500450386975
26.7555740779329 155.908382445121
27.0056261721192 156.808922503241
27.2556782663055 155.78568847365
27.5057303604918 157.129724519924
27.7557824546781 157.216920750904
28.0058345488643 157.642510427564
28.2558866430506 157.830304054436
28.5059387372369 158.092114777488
28.7559908314232 157.754385704631
29.0060429256095 158.047923137259
29.2560950197958 159.176621605502
29.5061471139821 160.848519045962
29.7561992081684 161.94844474718
30.0062513023547 161.161031015097
30.2563033965409 161.044402335526
30.5063554907272 161.23957152393
30.7564075849135 161.749540664506
31.0064596790998 161.905273295302
31.2565117732861 160.910691170493
31.5065638674724 157.619805410415
31.7566159616587 152.462820943745
32.006668055845 148.938034398345
32.2567201500313 125.506947415969
32.5067722442175 125.115755094085
32.7568243384038 127.17003895599
33.0068764325901 129.940328492012
33.2569285267764 130.169508428157
33.5069806209627 128.85415072575
33.757032715149 125.268891849252
34.0070848093353 120.429202771403
34.2571369035216 111.570530119292
34.5071889977079 98.6749179186618
34.7572410918941 87.0887610835501
35.0072931860804 79.8237661115635
35.2573452802667 84.0007265208923
35.507397374453 86.2154950214686
35.7574494686393 88.1459130000016
36.0075015628256 88.169019573673
36.2575536570119 88.2262735137494
36.5076057511982 89.0861768643152
36.7576578453845 90.0816743926099
37.0077099395707 90.4951407026398
37.257762033757 87.5103931333174
37.5078141279433 92.4223117629385
37.7578662221296 109.809747821702
38.0079183163159 114.820397111926
38.2579704105022 113.609797954169
38.5080225046885 112.53683150534
38.7580745988748 111.613670085003
39.0081266930611 98.580172533267
39.2581787872473 79.9106932067403
39.5082308814336 86.9837836351708
39.7582829756199 84.9833218576562
40.0083350698062 80.8405535968031
40.2583871639925 81.8414271743789
40.5084392581788 84.0621894276795
40.7584913523651 84.2375860000019
41.0085434465514 84.4883074693532
41.2585955407377 86.391272956321
41.5086476349239 90.37681213713
41.7586997291102 94.6552330832482
42.0087518232965 98.0148990383849
42.2588039174828 118.274160262462
42.5088560116691 117.769364907662
42.7589081058554 113.500942238108
43.0089602000417 110.689787991776
43.259012294228 110.690811980697
43.5090643884143 114.51275190225
43.7591164826005 118.592158650637
44.0091685767868 121.308480603969
44.2592206709731 120.724074267375
44.5092727651594 114.614867944182
44.7593248593457 108.319384200003
45.009376953532 105.781528307909
45.2594290477183 107.838859275879
45.5094811419046 114.093935788446
45.7595332360909 119.293558435926
46.0095853302771 122.294569671871
46.2596374244634 123.82589171923
46.5096895186497 122.684884029974
46.759741612836 123.001406827402
47.0097937070223 121.861470761693
47.2598458012086 120.812692635374
47.5098978953949 121.181440203934
47.7599499895812 122.453745116277
48.0100020837675 123.270029602011
48.2600541779537 123.497135664541
48.51010627214 127.969706036541
48.7601583663263 133.338795550615
49.0102104605126 136.749575737689
49.2602625546989 137.71187583419
49.5103146488852 135.958686374443
49.7603667430715 134.782383817193
50.0104188372578 133.202562232994
50.2604709314441 132.947891664398
50.5105230256303 132.789756107921
50.7605751198166 133.173822418196
51.0106272140029 133.745797079528
51.2606793081892 134.248820095115
51.5107314023755 137.287972902358
51.7607834965618 142.77712763566
52.0108355907481 147.655831615165
52.2608876849344 152.710657838091
52.5109397791207 160.356262717147
52.7609918733069 167.179407079269
53.0110439674932 171.765844851561
53.2610960616795 173.276605186376
53.5111481558658 164.305223809114
53.7612002500521 155.534492373756
54.0112523442384 150.879512507748
54.2613044384247 151.961921099606
54.511356532611 159.090760292717
54.7614086267972 165.46519441926
55.0114607209835 168.811459591353
55.2615128151698 169.89937022178
55.5115649093561 168.523082110094
55.7616170035424 166.684902884773
56.0116690977287 166.216154063738
56.261721191915 138.064843047261
56.5117732861013 137.341199093508
56.7618253802876 137.998909380881
57.0118774744738 140.530209093937
57.2619295686601 141.1041087014
57.5119816628464 141.444460792601
57.7620337570327 140.371744907461
58.012085851219 139.476933518381
58.2621379454053 135.754840588188
58.5121900395916 126.633910447188
58.7622421337779 117.440465484239
59.0122942279642 109.753203396566
59.2623463221504 105.578049606294
59.5123984163367 102.625292766745
59.762450510523 99.3891730655118
60.0125026047093 110.643978551412
60.2625546988956 124.093754018967
60.5126067930819 137.603317896041
60.7626588872682 151.175929507391
61.0127109814545 154.564147645366
61.2627630756408 152.850570074908
61.512815169827 151.123175228954
61.7628672640133 149.357746658842
62.0129193581996 139.46322111003
62.2629714523859 126.071595042048
62.5130235465722 113.52552039977
62.7630756407585 101.880212365814
63.0131277349448 94.1288578908444
63.2631798291311 104.93361205193
63.5132319233174 109.374248750423
63.7632840175037 114.001921454734
64.0133361116899 119.204180985293
64.2633882058762 119.658283537367
64.5134403000625 114.533399683793
64.7634923942488 102.419269345797
65.0135444884351 99.879761423487
65.2635965826214 109.803157908712
65.5136486768077 120.58697934833
65.763700770994 133.363546625372
66.0137528651803 145.790476450475
66.2638049593665 168.579471502398
66.5138570535528 167.650323094774
66.7639091477391 162.290846012353
67.0139612419254 158.794135984744
67.2640133361117 161.36402344348
67.514065430298 169.377856725744
67.7641175244843 177.365440420449
68.0141696186706 182.298762252303
68.2642217128569 182.858179041528
68.5142738070431 183.066028324281
68.7643259012294 182.371533505558
69.0143779954157 180.696582879247
69.264430089602 178.629104143887
69.5144821837883 177.27705161759
69.7645342779746 178.134692451808
70.0145863721609 179.167826616104
70.2646384663472 182.159682314109
70.5146905605335 182.910976392733
70.7647426547197 182.711743381214
71.014794748906 183.401785330305
71.2648468430923 183.65235521832
71.5148989372786 184.079232812967
71.7649510314649 184.148637360369
};
\addlegendentry{actual consumption}
\end{axis}

\end{tikzpicture}

%% file: self_con_eval.tex
\begin{tikzpicture}

\definecolor{crimson2143940}{RGB}{214,39,40}
\definecolor{darkgray176}{RGB}{176,176,176}
\definecolor{darkorange25512714}{RGB}{255,127,14}
\definecolor{darkturquoise23190207}{RGB}{23,190,207}
\definecolor{forestgreen4416044}{RGB}{44,160,44}
\definecolor{lightgray204}{RGB}{204,204,204}
\definecolor{steelblue31119180}{RGB}{31,119,180}

\begin{axis}[
width=250,
height=150,
legend cell align={left},
legend style={
  fill opacity=0.8,
  draw opacity=1,
  text opacity=1,
  at={(0.97,0.03)},
  anchor=south east,
  draw=lightgray204
},
tick align=outside,
tick pos=left,
x grid style={darkgray176},
xlabel={Percentage of Violations \(\displaystyle \Delta T_r\)},
xmin=-0.258583333333333, xmax=5.43025,
xtick style={color=black},
y grid style={darkgray176},
ylabel={Self Consumption \(\displaystyle \Delta S_r\)},
ymin=0.375858588432112, ymax=0.701267100733944,
ytick style={color=black}
]
\small
\addplot [semithick, steelblue31119180, opacity=0.7, mark=*, mark size=2, mark options={solid}, only marks]
table {%
0.0599305555555555 0.548169865284471
};
\addlegendentry{alpha 0.001}
\addplot [semithick, darkorange25512714, opacity=0.7, mark=triangle*, mark size=2, mark options={solid,rotate=180}, only marks]
table {%
0.268263888888889 0.602825217782798
};
\addlegendentry{alpha 0.5}
\addplot [semithick, forestgreen4416044, opacity=0.7, mark=pentagon*, mark size=2, mark options={solid}, only marks]
table {%
1.11465277777778 0.63129427167069
};
\addlegendentry{alpha 1.0}
\addplot [semithick, crimson2143940, opacity=0.7, mark=square*, mark size=2, mark options={solid}, only marks]
table {%
5.17166666666667 0.686475804720225
};
\addlegendentry{baseline}
\addplot [semithick, darkturquoise23190207, mark=diamond*, mark size=2, mark options={solid}, only marks]
table {%
0 0.390649884445832
};
\addlegendentry{no flexibility}
\end{axis}

\end{tikzpicture}

%% file: peak_red_3days.tex
% This file was created with tikzplotlib v0.10.1.
\begin{tikzpicture}

\definecolor{crimson2143940}{RGB}{214,39,40}
\definecolor{darkgray176}{RGB}{176,176,176}
\definecolor{darkorange25512714}{RGB}{255,127,14}
\definecolor{forestgreen4416044}{RGB}{44,160,44}
\definecolor{lightgray204}{RGB}{204,204,204}
\definecolor{steelblue31119180}{RGB}{31,119,180}

\begin{axis}[
width=250,
height=150,
legend cell align={left},
legend style={
  fill opacity=0.8,
  draw opacity=1,
  text opacity=1,
  at={(1,0)},%{(0.0,1)},
  anchor=south east,%north west,
  draw=lightgray204
},
tick align=outside,
tick pos=left,
x grid style={darkgray176},
xlabel={Time [h]},
xmajorgrids,
xmin=-3.58824755157324, xmax=75.3531985830381,
xtick style={color=black},
y grid style={darkgray176},
ylabel={Power [kW]},
ymajorgrids,
ymin=194.705524746642, ymax=396.63386076656,
ytick style={color=black}
]
\small
\addplot [line width=1.5pt, steelblue31119180, opacity=0.7]
table {%
0 250.90921087514
0.250052094186289 246.748160338551
0.500104188372578 240.872626688051
0.750156282558866 233.707878702767
1.00020837674516 229.691619236521
1.25026047093144 226.178704791032
1.50031256511773 223.29472297104
1.75036465930402 219.303997993421
2.00041675349031 219.880216723295
2.2504688476766 217.245899083116
2.50052094186289 215.994785648217
2.75057303604918 215.902306807383
3.00062513023547 215.265608637031
3.25067722442175 215.439931730176
3.50072931860804 214.521761249234
3.75078141279433 213.863966044458
4.00083350698062 215.70057478375
4.25088560116691 215.883141595695
4.5009376953532 217.267665561061
4.75098978953949 218.646566200843
5.00104188372578 218.91705963767
5.25109397791207 219.286936747087
5.50114607209835 219.199354407444
5.75119816628464 220.798081172912
6.00125026047093 224.375415212181
6.25130235465722 225.418350838677
6.50135444884351 226.562238548299
6.7514065430298 225.155106518845
7.00145863721609 219.723777335668
7.25151073140238 222.126524898088
7.50156282558866 224.700343970831
7.75161491977495 228.17305355108
8.00166701396124 224.893028943437
8.25171910814753 227.095153845775
8.50177120233382 230.928004913935
8.75182329652011 235.200584741299
9.0018753907064 236.478686313616
9.25192748489269 241.19386394385
9.50197957907898 241.899323527397
9.75203167326526 242.737065150903
10.0020837674516 241.355371975287
10.2521358616378 242.108916730675
10.5021879558241 249.098270513179
10.7522400500104 260.33612328321
11.0022921441967 261.752876172424
11.252344238383 265.9467956388
11.5023963325693 265.9467956388
11.7524484267556 265.9467956388
12.0025005209419 265.9467956388
12.2525526151282 265.9467956388
12.5026047093144 265.9467956388
12.7526568035007 265.9467956388
13.002708897687 265.9467956388
13.2527609918733 265.9467956388
13.5028130860596 265.9467956388
13.7528651802459 265.9467956388
14.0029172744322 264.25091423847
14.2529693686185 262.866600669639
14.5030214628048 260.184771666351
14.753073556991 259.993368199915
15.0031256511773 253.996588764091
15.2531777453636 255.366537987635
15.5032298395499 255.337522392693
15.7532819337362 254.244321121169
16.0033340279225 256.460542591385
16.2533861221088 261.388584134677
16.5034382162951 264.265278583758
16.7534903104814 265.9467956388
17.0035424046676 265.9467956388
17.2535944988539 265.9467956388
17.5036465930402 265.9467956388
17.7536986872265 265.9467956388
18.0037507814128 265.9467956388
18.2538028755991 265.9467956388
18.5038549697854 265.9467956388
18.7539070639717 265.9467956388
19.003959158158 265.9467956388
19.2540112523442 265.9467956388
19.5040633465305 265.9467956388
19.7541154407168 265.9467956388
20.0041675349031 265.9467956388
20.2542196290894 265.9467956388
20.5042717232757 265.9467956388
20.754323817462 265.9467956388
21.0043759116483 265.9467956388
21.2544280058346 265.9467956388
21.5044801000208 265.9467956388
21.7545321942071 263.099146222245
22.0045842883934 259.047425376287
22.2546363825797 255.051344850491
22.504688476766 250.346495777454
22.7547405709523 246.617637978707
23.0047926651386 265.9467956388
23.2548447593249 262.34986677844
23.5048968535111 258.087003445624
23.7549489476974 250.779912640593
24.0050010418837 251.101589816422
24.25505313607 247.658125354608
24.5051052302563 240.894660943405
24.7551573244426 235.717261251625
25.0052094186289 230.762040301337
25.2552615128152 226.24245494129
25.5053136070015 220.720674374804
25.7553657011877 218.610358869778
26.005417795374 218.017231909233
26.2554698895603 216.543595447601
26.5055219837466 214.453344121279
26.7555740779329 212.295651141849
27.0056261721192 214.040335722084
27.2556782663055 213.695285403856
27.5057303604918 213.34507080703
27.7557824546781 214.277242221062
28.0058345488643 217.762212181244
28.2558866430506 219.317099959496
28.5059387372369 222.72242526751
28.7559908314232 224.608247669153
29.0060429256095 231.66293913941
29.2560950197958 233.623949612059
29.5061471139821 241.068544116521
29.7561992081684 248.501471176741
30.0062513023547 261.363757343488
30.2563033965409 269.400503333648
30.5063554907272 278.723192948813
30.7564075849135 288.389460106456
31.0064596790998 291.163959220155
31.2565117732861 301.497527680588
31.5065638674724 311.394312933313
31.7566159616587 319.562441996418
32.006668055845 322.251188234098
32.2567201500313 322.251188234098
32.5067722442175 322.251188234098
32.7568243384038 322.251188234098
33.0068764325901 322.251188234098
33.2569285267764 322.251188234098
33.5069806209627 322.251188234098
33.757032715149 322.251188234098
34.0070848093353 322.251188234098
34.2571369035216 322.251188234098
34.5071889977079 322.251188234098
34.7572410918941 322.251188234098
35.0072931860804 322.251188234098
35.2573452802667 322.251188234098
35.507397374453 322.251188234098
35.7574494686393 322.251188234098
36.0075015628256 322.251188234098
36.2575536570119 322.251188234098
36.5076057511982 322.251188234098
36.7576578453845 322.251188234098
37.0077099395707 322.251188234098
37.257762033757 322.251188234098
37.5078141279433 322.251188234098
37.7578662221296 322.251188234098
38.0079183163159 322.251188234098
38.2579704105022 322.251188234098
38.5080225046885 322.251188234098
38.7580745988748 322.251188234098
39.0081266930611 322.251188234098
39.2581787872473 322.251188234098
39.5082308814336 322.251188234098
39.7582829756199 322.251188234098
40.0083350698062 322.251188234098
40.2583871639925 322.251188234098
40.5084392581788 322.251188234098
40.7584913523651 322.251188234098
41.0085434465514 322.251188234098
41.2585955407377 322.251188234098
41.5086476349239 322.251188234098
41.7586997291102 322.251188234098
42.0087518232965 322.251188234098
42.2588039174828 322.251188234098
42.5088560116691 322.251188234098
42.7589081058554 322.251188234098
43.0089602000417 322.251188234098
43.259012294228 322.251188234098
43.5090643884143 322.251188234098
43.7591164826005 322.251188234098
44.0091685767868 322.251188234098
44.2592206709731 319.844373353994
44.5092727651594 312.707282949651
44.7593248593457 307.090384022159
45.009376953532 301.805184796369
45.2594290477183 296.20913425285
45.5094811419046 290.537092306098
45.7595332360909 284.922259090046
46.0095853302771 281.850048878608
46.2596374244634 276.914280044376
46.5096895186497 271.006517470028
46.759741612836 264.696897512866
47.0097937070223 308.887705990185
47.2598458012086 305.411897842777
47.5098978953949 297.28209920305
47.7599499895812 288.53063618178
48.0100020837675 284.123134794284
48.2600541779537 275.49144478259
48.51010627214 267.863499260914
48.7601583663263 260.266022982972
49.0102104605126 254.987441684571
49.2602625546989 248.599802866653
49.5103146488852 241.396188903175
49.7603667430715 240.694822488591
50.0104188372578 237.444406411581
50.2604709314441 236.906403343339
50.5105230256303 233.797909293005
50.7605751198166 229.280427898314
51.0106272140029 231.47740764901
51.2606793081892 232.66676013726
51.5107314023755 232.391963134933
51.7607834965618 231.812798853562
52.0108355907481 234.462838853031
52.2608876849344 236.743958090971
52.5109397791207 239.06065347037
52.7609918733069 241.503567644113
53.0110439674932 246.540078164937
53.2610960616795 250.694300377601
53.5111481558658 258.609379789366
53.7612002500521 267.647039479796
54.0112523442384 281.404754174692
54.2613044384247 287.718677697352
54.511356532611 298.311798974767
54.7614086267972 303.852649118152
55.0114607209835 308.985884664435
55.2615128151698 316.930190067678
55.5115649093561 325.547993908029
55.7616170035424 335.730344640307
56.0116690977287 338.557360701735
56.261721191915 338.557360701735
56.5117732861013 338.557360701735
56.7618253802876 338.557360701735
57.0118774744738 338.557360701735
57.2619295686601 338.557360701735
57.5119816628464 338.557360701735
57.7620337570327 338.557360701735
58.012085851219 338.557360701735
58.2621379454053 338.557360701735
58.5121900395916 338.557360701735
58.7622421337779 338.557360701735
59.0122942279642 338.557360701735
59.2623463221504 338.557360701735
59.5123984163367 338.557360701735
59.762450510523 338.557360701735
60.0125026047093 338.557360701735
60.2625546988956 338.557360701735
60.5126067930819 338.557360701735
60.7626588872682 338.557360701735
61.0127109814545 338.557360701735
61.2627630756408 338.557360701735
61.512815169827 338.557360701735
61.7628672640133 338.557360701735
62.0129193581996 338.557360701735
62.2629714523859 338.557360701735
62.5130235465722 338.557360701735
62.7630756407585 338.557360701735
63.0131277349448 338.557360701735
63.2631798291311 338.557360701735
63.5132319233174 338.557360701735
63.7632840175037 338.557360701735
64.0133361116899 338.557360701735
64.2633882058762 338.557360701735
64.5134403000625 338.557360701735
64.7634923942488 338.557360701735
65.0135444884351 338.557360701735
65.2635965826214 338.557360701735
65.5136486768077 338.557360701735
65.763700770994 338.557360701735
66.0137528651803 338.557360701735
66.2638049593665 338.557360701735
66.5138570535528 338.557360701735
66.7639091477391 338.557360701735
67.0139612419254 338.557360701735
67.2640133361117 338.557360701735
67.514065430298 338.557360701735
67.7641175244843 338.557360701735
68.0141696186706 335.198997753369
68.2642217128569 328.296327343494
68.5142738070431 322.091637876676
68.7643259012294 319.932548629328
69.0143779954157 312.425944527765
69.264430089602 308.503770566805
69.5144821837883 302.337507158153
69.7645342779746 299.951285287249
70.0145863721609 293.29447279259
70.2646384663472 286.477280757002
70.5146905605335 282.376920057427
70.7647426547197 275.998194401831
71.014794748906 321.220672727518
71.2648468430923 318.69465192225
71.5148989372786 310.830909265148
71.7649510314649 300.296641636801
};
\addlegendentry{baseline + request}
\addplot [line width=1.5pt, crimson2143940, opacity=0.7, dash pattern=on 1pt off 3pt on 3pt off 3pt]
table {%
0 250.90921087514
0.250052094186289 246.748160338551
0.500104188372578 240.872626688051
0.750156282558866 233.707878702767
1.00020837674516 229.691619236521
1.25026047093144 226.178704791032
1.50031256511773 223.29472297104
1.75036465930402 219.303997993421
2.00041675349031 219.880216723295
2.2504688476766 217.245899083116
2.50052094186289 215.994785648217
2.75057303604918 215.902306807383
3.00062513023547 215.265608637031
3.25067722442175 215.439931730176
3.50072931860804 214.521761249234
3.75078141279433 213.863966044458
4.00083350698062 215.70057478375
4.25088560116691 215.883141595695
4.5009376953532 217.267665561061
4.75098978953949 218.646566200843
5.00104188372578 218.91705963767
5.25109397791207 219.286936747087
5.50114607209835 219.199354407444
5.75119816628464 220.798081172912
6.00125026047093 224.375415212181
6.25130235465722 225.418350838677
6.50135444884351 226.562238548299
6.7514065430298 225.155106518845
7.00145863721609 219.723777335668
7.25151073140238 222.126524898088
7.50156282558866 224.700343970831
7.75161491977495 228.17305355108
8.00166701396124 224.893028943437
8.25171910814753 227.095153845775
8.50177120233382 230.928004913935
8.75182329652011 235.200584741299
9.0018753907064 236.478686313616
9.25192748489269 241.19386394385
9.50197957907898 241.899323527397
9.75203167326526 242.737065150903
10.0020837674516 241.355371975287
10.2521358616378 242.108916730675
10.5021879558241 249.098270513179
10.7522400500104 260.33612328321
11.0022921441967 261.752876172424
11.252344238383 265.9467956388
11.5023963325693 265.9467956388
11.7524484267556 265.9467956388
12.0025005209419 265.9467956388
12.2525526151282 265.9467956388
12.5026047093144 265.9467956388
12.7526568035007 265.9467956388
13.002708897687 265.9467956388
13.2527609918733 265.9467956388
13.5028130860596 265.9467956388
13.7528651802459 265.9467956388
14.0029172744322 264.25091423847
14.2529693686185 262.866600669639
14.5030214628048 260.184771666351
14.753073556991 259.993368199915
15.0031256511773 253.996588764091
15.2531777453636 255.366537987635
15.5032298395499 255.337522392693
15.7532819337362 254.244321121169
16.0033340279225 256.460542591385
16.2533861221088 261.388584134677
16.5034382162951 264.265278583758
16.7534903104814 265.9467956388
17.0035424046676 265.9467956388
17.2535944988539 265.9467956388
17.5036465930402 265.9467956388
17.7536986872265 265.9467956388
18.0037507814128 265.9467956388
18.2538028755991 265.9467956388
18.5038549697854 265.9467956388
18.7539070639717 265.9467956388
19.003959158158 265.9467956388
19.2540112523442 265.9467956388
19.5040633465305 265.9467956388
19.7541154407168 265.9467956388
20.0041675349031 265.9467956388
20.2542196290894 265.9467956388
20.5042717232757 265.9467956388
20.754323817462 265.9467956388
21.0043759116483 265.9467956388
21.2544280058346 265.9467956388
21.5044801000208 265.9467956388
21.7545321942071 263.099146222245
22.0045842883934 259.047425376287
22.2546363825797 255.051344850491
22.504688476766 250.346495777454
22.7547405709523 246.617637978707
23.0047926651386 265.9467956388
23.2548447593249 262.34986677844
23.5048968535111 258.087003445624
23.7549489476974 250.779912640593
24.0050010418837 251.101589816422
24.25505313607 247.658125354608
24.5051052302563 240.894660943405
24.7551573244426 235.717261251625
25.0052094186289 230.762040301337
25.2552615128152 226.24245494129
25.5053136070015 220.720674374804
25.7553657011877 218.610358869778
26.005417795374 218.017231909233
26.2554698895603 216.543595447601
26.5055219837466 214.453344121279
26.7555740779329 212.295651141849
27.0056261721192 214.040335722084
27.2556782663055 213.695285403856
27.5057303604918 213.34507080703
27.7557824546781 214.277242221062
28.0058345488643 217.762212181244
28.2558866430506 219.317099959496
28.5059387372369 222.72242526751
28.7559908314232 224.608247669153
29.0060429256095 231.66293913941
29.2560950197958 233.623949612059
29.5061471139821 241.068544116521
29.7561992081684 248.501471176741
30.0062513023547 261.363757343488
30.2563033965409 269.400503333648
30.5063554907272 278.723192948813
30.7564075849135 288.389460106456
31.0064596790998 291.163959220155
31.2565117732861 301.497527680588
31.5065638674724 311.394312933313
31.7566159616587 319.562441996418
32.006668055845 326.112086146674
32.2567201500313 331.100473241741
32.5067722442175 334.055473455944
32.7568243384038 335.513387234098
33.0068764325901 335.513387234098
33.2569285267764 335.513387234098
33.5069806209627 335.513387234098
33.757032715149 335.513387234098
34.0070848093353 335.513387234098
34.2571369035216 335.513387234098
34.5071889977079 335.513387234098
34.7572410918941 335.513387234098
35.0072931860804 335.513387234098
35.2573452802667 335.513387234098
35.507397374453 335.513387234098
35.7574494686393 335.513387234098
36.0075015628256 335.513387234098
36.2575536570119 335.513387234098
36.5076057511982 335.513387234098
36.7576578453845 335.513387234098
37.0077099395707 335.513387234098
37.257762033757 335.513387234098
37.5078141279433 335.513387234098
37.7578662221296 335.513387234098
38.0079183163159 335.513387234098
38.2579704105022 335.513387234098
38.5080225046885 335.513387234098
38.7580745988748 335.513387234098
39.0081266930611 335.513387234098
39.2581787872473 335.513387234098
39.5082308814336 335.513387234098
39.7582829756199 335.513387234098
40.0083350698062 335.513387234098
40.2583871639925 335.513387234098
40.5084392581788 335.513387234098
40.7584913523651 335.513387234098
41.0085434465514 335.513387234098
41.2585955407377 335.513387234098
41.5086476349239 335.513387234098
41.7586997291102 335.513387234098
42.0087518232965 335.513387234098
42.2588039174828 335.513387234098
42.5088560116691 335.513387234098
42.7589081058554 335.513387234098
43.0089602000417 335.513387234098
43.259012294228 335.513387234098
43.5090643884143 335.513387234098
43.7591164826005 335.014193087365
44.0091685767868 327.087752262682
44.2592206709731 319.844373353994
44.5092727651594 312.707282949651
44.7593248593457 307.090384022159
45.009376953532 301.805184796369
45.2594290477183 296.20913425285
45.5094811419046 290.537092306098
45.7595332360909 284.922259090046
46.0095853302771 281.850048878608
46.2596374244634 276.914280044376
46.5096895186497 271.006517470028
46.759741612836 264.696897512866
47.0097937070223 308.887705990185
47.2598458012086 305.411897842777
47.5098978953949 297.28209920305
47.7599499895812 288.53063618178
48.0100020837675 284.123134794284
48.2600541779537 275.49144478259
48.51010627214 267.863499260914
48.7601583663263 260.266022982972
49.0102104605126 254.987441684571
49.2602625546989 248.599802866653
49.5103146488852 241.396188903175
49.7603667430715 240.694822488591
50.0104188372578 237.444406411581
50.2604709314441 236.906403343339
50.5105230256303 233.797909293005
50.7605751198166 229.280427898314
51.0106272140029 231.47740764901
51.2606793081892 232.66676013726
51.5107314023755 232.391963134933
51.7607834965618 231.812798853562
52.0108355907481 234.462838853031
52.2608876849344 236.743958090971
52.5109397791207 239.06065347037
52.7609918733069 241.503567644113
53.0110439674932 246.540078164937
53.2610960616795 250.694300377601
53.5111481558658 258.609379789366
53.7612002500521 267.647039479796
54.0112523442384 281.404754174692
54.2613044384247 287.718677697352
54.511356532611 298.311798974767
54.7614086267972 303.852649118152
55.0114607209835 308.985884664435
55.2615128151698 316.930190067678
55.5115649093561 325.547993908029
55.7616170035424 335.730344640307
56.0116690977287 343.25412474995
56.261721191915 346.527138101735
56.5117732861013 346.527138101735
56.7618253802876 346.527138101735
57.0118774744738 346.527138101735
57.2619295686601 346.527138101735
57.5119816628464 346.527138101735
57.7620337570327 346.527138101735
58.012085851219 346.527138101735
58.2621379454053 346.527138101735
58.5121900395916 346.527138101735
58.7622421337779 346.527138101735
59.0122942279642 346.527138101735
59.2623463221504 346.527138101735
59.5123984163367 346.527138101735
59.762450510523 346.527138101735
60.0125026047093 346.527138101735
60.2625546988956 346.527138101735
60.5126067930819 346.527138101735
60.7626588872682 346.527138101735
61.0127109814545 346.527138101735
61.2627630756408 346.527138101735
61.512815169827 346.527138101735
61.7628672640133 346.527138101735
62.0129193581996 346.527138101735
62.2629714523859 346.527138101735
62.5130235465722 346.527138101735
62.7630756407585 346.527138101735
63.0131277349448 346.527138101735
63.2631798291311 346.527138101735
63.5132319233174 346.527138101735
63.7632840175037 346.527138101735
64.0133361116899 346.527138101735
64.2633882058762 346.527138101735
64.5134403000625 346.527138101735
64.7634923942488 346.527138101735
65.0135444884351 346.527138101735
65.2635965826214 346.527138101735
65.5136486768077 346.527138101735
65.763700770994 346.527138101735
66.0137528651803 346.527138101735
66.2638049593665 346.527138101735
66.5138570535528 346.527138101735
66.7639091477391 346.527138101735
67.0139612419254 346.527138101735
67.2640133361117 346.527138101735
67.514065430298 346.527138101735
67.7641175244843 342.679168399289
68.0141696186706 335.198997753369
68.2642217128569 328.296327343494
68.5142738070431 322.091637876676
68.7643259012294 319.932548629328
69.0143779954157 312.425944527765
69.264430089602 308.503770566805
69.5144821837883 302.337507158153
69.7645342779746 299.951285287249
70.0145863721609 293.29447279259
70.2646384663472 286.477280757002
70.5146905605335 282.376920057427
70.7647426547197 275.998194401831
71.014794748906 321.220672727518
71.2648468430923 318.69465192225
71.5148989372786 310.830909265148
71.7649510314649 300.296641636801
};
\addlegendentry{committed trajectory}
\addplot [line width=1.5pt, forestgreen4416044, opacity=0.7, dashed]
table {%
0 250.90921087514
0.250052094186289 246.748160338551
0.500104188372578 240.872626688051
0.750156282558866 233.707878702767
1.00020837674516 229.691619236521
1.25026047093144 226.178704791032
1.50031256511773 223.29472297104
1.75036465930402 219.303997993421
2.00041675349031 219.880216723295
2.2504688476766 217.245899083116
2.50052094186289 215.994785648217
2.75057303604918 215.902306807383
3.00062513023547 215.265608637031
3.25067722442175 215.439931730176
3.50072931860804 214.521761249234
3.75078141279433 213.863966044458
4.00083350698062 215.70057478375
4.25088560116691 215.883141595695
4.5009376953532 217.267665561061
4.75098978953949 218.646566200843
5.00104188372578 218.91705963767
5.25109397791207 219.286936747087
5.50114607209835 219.199354407444
5.75119816628464 220.798081172912
6.00125026047093 224.375415212181
6.25130235465722 225.418350838677
6.50135444884351 226.562238548299
6.7514065430298 225.155106518845
7.00145863721609 219.723777335668
7.25151073140238 222.126524898088
7.50156282558866 224.700343970831
7.75161491977495 228.17305355108
8.00166701396124 224.893028943437
8.25171910814753 227.095153845775
8.50177120233382 230.928004913935
8.75182329652011 235.200584741299
9.0018753907064 236.478686313616
9.25192748489269 241.19386394385
9.50197957907898 241.899323527397
9.75203167326526 242.737065150903
10.0020837674516 241.355371975287
10.2521358616378 242.108916730675
10.5021879558241 249.098270513179
10.7522400500104 260.33612328321
11.0022921441967 261.752876172424
11.252344238383 267.172939577473
11.5023963325693 275.77720153172
11.7524484267556 281.003757504978
12.0025005209419 275.959921359327
12.2525526151282 271.148860289776
12.5026047093144 278.704219006702
12.7526568035007 280.227183892431
13.002708897687 277.1157442906
13.2527609918733 275.117914424238
13.5028130860596 269.855877944311
13.7528651802459 266.538804419462
14.0029172744322 264.25091423847
14.2529693686185 262.866600669639
14.5030214628048 260.184771666351
14.753073556991 259.993368199915
15.0031256511773 253.996588764091
15.2531777453636 255.366537987635
15.5032298395499 255.337522392693
15.7532819337362 254.244321121169
16.0033340279225 256.460542591385
16.2533861221088 261.388584134677
16.5034382162951 264.265278583758
16.7534903104814 271.673225837126
17.0035424046676 279.527271126638
17.2535944988539 292.081307241913
17.5036465930402 298.436468485532
17.7536986872265 303.479692568534
18.0037507814128 307.201817678145
18.2538028755991 310.854760581491
18.5038549697854 312.145141660698
18.7539070639717 315.41305890407
19.003959158158 311.482449954732
19.2540112523442 307.163162096486
19.5040633465305 302.185466970814
19.7541154407168 298.395479415024
20.0041675349031 291.597777749408
20.2542196290894 286.354160529817
20.5042717232757 281.465424884748
20.754323817462 279.33502607406
21.0043759116483 274.756797461481
21.2544280058346 269.098565178242
21.5044801000208 267.222344509401
21.7545321942071 263.099146222245
22.0045842883934 259.047425376287
22.2546363825797 255.051344850491
22.504688476766 250.346495777454
22.7547405709523 246.617637978707
23.0047926651386 267.077056138156
23.2548447593249 262.34986677844
23.5048968535111 258.087003445624
23.7549489476974 250.779912640593
24.0050010418837 251.101589816422
24.25505313607 247.658125354608
24.5051052302563 240.894660943405
24.7551573244426 235.717261251625
25.0052094186289 230.762040301337
25.2552615128152 226.24245494129
25.5053136070015 220.720674374804
25.7553657011877 218.610358869778
26.005417795374 218.017231909233
26.2554698895603 216.543595447601
26.5055219837466 214.453344121279
26.7555740779329 212.295651141849
27.0056261721192 214.040335722084
27.2556782663055 213.695285403856
27.5057303604918 213.34507080703
27.7557824546781 214.277242221062
28.0058345488643 217.762212181244
28.2558866430506 219.317099959496
28.5059387372369 222.72242526751
28.7559908314232 224.608247669153
29.0060429256095 231.66293913941
29.2560950197958 233.623949612059
29.5061471139821 241.068544116521
29.7561992081684 248.501471176741
30.0062513023547 261.363757343488
30.2563033965409 269.400503333648
30.5063554907272 278.723192948813
30.7564075849135 288.389460106456
31.0064596790998 291.163959220155
31.2565117732861 301.497527680588
31.5065638674724 311.394312933313
31.7566159616587 319.562441996418
32.006668055845 326.112086146674
32.2567201500313 331.100473241741
32.5067722442175 334.055473455944
32.7568243384038 338.209447018158
33.0068764325901 343.021234912104
33.2569285267764 346.318033861714
33.5069806209627 348.683502983434
33.757032715149 351.50646980813
34.0070848093353 354.381825370166
34.2571369035216 358.592605996965
34.5071889977079 361.109522370333
34.7572410918941 362.865453601738
35.0072931860804 364.094322384747
35.2573452802667 366.177075943622
35.507397374453 367.742254025156
35.7574494686393 371.641150478507
36.0075015628256 372.11421752511
36.2575536570119 369.065265694324
36.5076057511982 364.883997963369
36.7576578453845 360.880171019675
37.0077099395707 360.688442394957
37.257762033757 355.906416047305
37.5078141279433 356.7633228825
37.7578662221296 358.462255779741
38.0079183163159 355.953736140836
38.2579704105022 349.68581295162
38.5080225046885 345.860497904823
38.7580745988748 345.735369347071
39.0081266930611 342.614978329003
39.2581787872473 339.584177981144
39.5082308814336 342.12001091575
39.7582829756199 342.346321080121
40.0083350698062 342.434152070216
40.2583871639925 342.039945470511
40.5084392581788 341.142967658782
40.7584913523651 345.215364746093
41.0085434465514 350.584320919971
41.2585955407377 357.938712691575
41.5086476349239 360.512608272149
41.7586997291102 364.884648279933
42.0087518232965 367.721596910761
42.2588039174828 370.74260425624
42.5088560116691 373.006930025256
42.7589081058554 372.149851047441
43.0089602000417 367.989106542176
43.259012294228 354.91500844505
43.5090643884143 344.477331953458
43.7591164826005 335.014193087365
44.0091685767868 327.087752262682
44.2592206709731 319.844373353994
44.5092727651594 312.707282949651
44.7593248593457 307.090384022159
45.009376953532 301.805184796369
45.2594290477183 296.20913425285
45.5094811419046 290.537092306098
45.7595332360909 284.922259090046
46.0095853302771 281.850048878608
46.2596374244634 276.914280044376
46.5096895186497 271.006517470028
46.759741612836 264.696897512866
47.0097937070223 308.887705990185
47.2598458012086 305.411897842777
47.5098978953949 297.28209920305
47.7599499895812 288.53063618178
48.0100020837675 284.123134794284
48.2600541779537 275.49144478259
48.51010627214 267.863499260914
48.7601583663263 260.266022982972
49.0102104605126 254.987441684571
49.2602625546989 248.599802866653
49.5103146488852 241.396188903175
49.7603667430715 240.694822488591
50.0104188372578 237.444406411581
50.2604709314441 236.906403343339
50.5105230256303 233.797909293005
50.7605751198166 229.280427898314
51.0106272140029 231.47740764901
51.2606793081892 232.66676013726
51.5107314023755 232.391963134933
51.7607834965618 231.812798853562
52.0108355907481 234.462838853031
52.2608876849344 236.743958090971
52.5109397791207 239.06065347037
52.7609918733069 241.503567644113
53.0110439674932 246.540078164937
53.2610960616795 250.694300377601
53.5111481558658 258.609379789366
53.7612002500521 267.647039479796
54.0112523442384 281.404754174692
54.2613044384247 287.718677697352
54.511356532611 298.311798974767
54.7614086267972 303.852649118152
55.0114607209835 308.985884664435
55.2615128151698 316.930190067678
55.5115649093561 325.547993908029
55.7616170035424 335.730344640307
56.0116690977287 343.25412474995
56.261721191915 348.617476725201
56.5117732861013 351.286949713767
56.7618253802876 355.322374391702
57.0118774744738 357.039267502293
57.2619295686601 361.389541032132
57.5119816628464 362.944620079961
57.7620337570327 365.479152381438
58.012085851219 368.065576119353
58.2621379454053 372.088396132116
58.5121900395916 374.538502041982
58.7622421337779 371.217067570761
59.0122942279642 371.236959606845
59.2623463221504 375.597446932134
59.5123984163367 378.362382566946
59.762450510523 387.455300038382
60.0125026047093 385.258817588588
60.2625546988956 379.997450552183
60.5126067930819 376.531358899323
60.7626588872682 373.439180143971
61.0127109814545 375.005238065562
61.2627630756408 374.977350961129
61.512815169827 373.596403736866
61.7628672640133 370.680384262604
62.0129193581996 370.236983127507
62.2629714523859 371.546012990543
62.5130235465722 367.240324399273
62.7630756407585 366.701173713564
63.0131277349448 367.381672617217
63.2631798291311 366.369799170145
63.5132319233174 367.393129664923
63.7632840175037 364.719582634351
64.0133361116899 358.89301399575
64.2633882058762 358.722076374132
64.5134403000625 359.353036457152
64.7634923942488 362.457858131593
65.0135444884351 362.353539703898
65.2635965826214 368.984645567311
65.5136486768077 373.878258586611
65.763700770994 379.295644217571
66.0137528651803 378.223903392669
66.2638049593665 380.064337523516
66.5138570535528 380.533120131578
66.7639091477391 381.264611504654
67.0139612419254 371.850246380883
67.2640133361117 360.08271076924
67.514065430298 349.337358034909
67.7641175244843 342.679168399289
68.0141696186706 335.198997753369
68.2642217128569 328.296327343494
68.5142738070431 322.091637876676
68.7643259012294 319.932548629328
69.0143779954157 312.425944527765
69.264430089602 308.503770566805
69.5144821837883 302.337507158153
69.7645342779746 299.951285287249
70.0145863721609 293.29447279259
70.2646384663472 286.477280757002
70.5146905605335 282.376920057427
70.7647426547197 275.998194401831
71.014794748906 321.220672727518
71.2648468430923 318.69465192225
71.5148989372786 310.830909265148
71.7649510314649 300.296641636801
};
\addlegendentry{baseline consumption}
\addplot [line width=1.5pt, darkorange25512714, opacity=0.7, dotted]
table {%
0 252.084517742652
0.250052094186289 249.541295872474
0.500104188372578 249.813527966765
0.750156282558866 245.683942643147
1.00020837674516 237.933261338921
1.25026047093144 231.919385622727
1.50031256511773 224.562714173559
1.75036465930402 218.304922465472
2.00041675349031 217.239873464667
2.2504688476766 213.914765795847
2.50052094186289 211.855457435811
2.75057303604918 212.289039226425
3.00062513023547 213.032418833576
3.25067722442175 213.258410331654
3.50072931860804 214.418594692971
3.75078141279433 214.979201318694
4.00083350698062 218.824496081539
4.25088560116691 219.610910128966
4.5009376953532 219.859387263088
4.75098978953949 220.172346084098
5.00104188372578 220.361435739264
5.25109397791207 221.461381519527
5.50114607209835 221.743514200757
5.75119816628464 222.683933287947
6.00125026047093 225.904721483393
6.25130235465722 226.858021817774
6.50135444884351 228.34800183682
6.7514065430298 227.471169434738
7.00145863721609 221.792168302608
7.25151073140238 223.950662405817
7.50156282558866 226.480353586739
7.75161491977495 230.048005317104
8.00166701396124 211.002257376895
8.25171910814753 203.88408547482
8.50177120233382 207.765053008762
8.75182329652011 213.572265106476
9.0018753907064 218.252867530707
9.25192748489269 226.526107328674
9.50197957907898 224.963466758384
9.75203167326526 225.566704565021
10.0020837674516 228.787826159484
10.2521358616378 234.451085398452
10.5021879558241 246.931017740308
10.7522400500104 262.765602136592
11.0022921441967 266.755704501613
11.252344238383 265.9467956388
11.5023963325693 265.9467956388
11.7524484267556 265.9467956388
12.0025005209419 265.9467956388
12.2525526151282 265.9467956388
12.5026047093144 265.9467956388
12.7526568035007 265.9467956388
13.002708897687 265.9467956388
13.2527609918733 265.9467956388
13.5028130860596 265.9467956388
13.7528651802459 265.9467956388
14.0029172744322 264.624115403731
14.2529693686185 263.541371711183
14.5030214628048 261.453063759933
14.753073556991 261.309656928042
15.0031256511773 256.741394812666
15.2531777453636 258.061833167574
15.5032298395499 258.371299132581
15.7532819337362 258.011057152508
16.0033340279225 259.794434458207
16.2533861221088 263.025114631323
16.5034382162951 264.879424648118
16.7534903104814 265.9467956388
17.0035424046676 265.9467956388
17.2535944988539 265.9467956388
17.5036465930402 265.9467956388
17.7536986872265 265.9467956388
18.0037507814128 265.9467956388
18.2538028755991 265.9467956388
18.5038549697854 265.9467956388
18.7539070639717 265.9467956388
19.003959158158 265.9467956388
19.2540112523442 265.9467956388
19.5040633465305 265.9467956388
19.7541154407168 265.9467956388
20.0041675349031 265.9467956388
20.2542196290894 265.9467956388
20.5042717232757 265.9467956388
20.754323817462 265.9467956388
21.0043759116483 265.9467956388
21.2544280058346 265.9467956388
21.5044801000208 265.9467956388
21.7545321942071 264.880658335641
22.0045842883934 263.431236858258
22.2546363825797 262.082704027699
22.504688476766 260.603325060861
22.7547405709523 259.518255283887
23.0047926651386 265.9467956388
23.2548447593249 264.643272057973
23.5048968535111 263.261959501757
23.7549489476974 261.215572238991
24.0050010418837 286.130163335729
24.25505313607 282.686698873915
24.5051052302563 274.539165885612
24.7551573244426 260.91186140962
25.0052094186289 251.038677494261
25.2552615128152 245.231347291683
25.5053136070015 237.795559711596
25.7553657011877 236.092803964113
26.005417795374 236.204480669332
26.2554698895603 234.519648358613
26.5055219837466 231.945070520274
26.7555740779329 229.2841854524
27.0056261721192 230.63841800986
27.2556782663055 230.84809196098
27.5057303604918 230.678192130545
27.7557824546781 231.08271048484
28.0058345488643 234.031985284158
28.2558866430506 233.508752565023
28.5059387372369 235.562015380125
28.7559908314232 236.830270015401
29.0060429256095 243.718634845188
29.2560950197958 246.69408278629
29.5061471139821 254.494129870138
29.7561992081684 261.835429342987
30.0062513023547 274.244899844395
30.2563033965409 280.574533971895
30.5063554907272 288.829606486106
30.7564075849135 297.550891084694
31.0064596790998 299.79656897612
31.2565117732861 309.609411939222
31.5065638674724 316.545770577814
31.7566159616587 320.687411266435
32.006668055845 326.112086146674
32.2567201500313 331.100473241741
32.5067722442175 334.055473455944
32.7568243384038 335.513387234098
33.0068764325901 335.513387234098
33.2569285267764 335.513387234098
33.5069806209627 335.513387234098
33.757032715149 335.513387234098
34.0070848093353 335.513387234098
34.2571369035216 335.513387234098
34.5071889977079 335.513387234098
34.7572410918941 335.513387234098
35.0072931860804 335.513387234098
35.2573452802667 335.513387234098
35.507397374453 335.513387234098
35.7574494686393 335.513387234098
36.0075015628256 335.513387234098
36.2575536570119 335.513387234098
36.5076057511982 335.513387234098
36.7576578453845 335.513387234098
37.0077099395707 335.513387234098
37.257762033757 335.513387234098
37.5078141279433 335.513387234098
37.7578662221296 335.513387234098
38.0079183163159 335.513387234098
38.2579704105022 335.513387234098
38.5080225046885 335.513387234098
38.7580745988748 335.513387234098
39.0081266930611 335.513387234098
39.2581787872473 335.513387234098
39.5082308814336 335.513387234098
39.7582829756199 335.513387234098
40.0083350698062 335.513387234098
40.2583871639925 335.513387234098
40.5084392581788 335.513387234098
40.7584913523651 335.513387234098
41.0085434465514 335.513387234098
41.2585955407377 335.513387234098
41.5086476349239 335.513387234098
41.7586997291102 335.513387234099
42.0087518232965 335.513387234098
42.2588039174828 335.513387234098
42.5088560116691 335.513387234098
42.7589081058554 335.513387234098
43.0089602000417 335.513387234098
43.259012294228 335.513387234098
43.5090643884143 335.513387234098
43.7591164826005 335.014193087366
44.0091685767868 327.087752262682
44.2592206709731 325.626943075952
44.5092727651594 321.122970597143
44.7593248593457 317.528206535365
45.009376953532 314.09934759478
45.2594290477183 311.173119135559
45.5094811419046 308.395586119054
45.7595332360909 305.800667122217
46.0095853302771 304.687765346139
46.2596374244634 302.289859829945
46.5096895186497 299.573393751426
46.759741612836 296.816716390504
47.0097937070223 319.269553525209
47.2598458012086 317.277693455618
47.5098978953949 312.754300360483
47.7599499895812 307.746526254218
48.0100020837675 306.52668172854
48.2600541779537 297.756272966926
48.51010627214 291.248727959786
48.7601583663263 284.711717146355
49.0102104605126 278.940537859287
49.2602625546989 269.632989968636
49.5103146488852 261.100467507578
49.7603667430715 258.218727935567
50.0104188372578 254.160402288206
50.2604709314441 253.878752328893
50.5105230256303 248.631094750586
50.7605751198166 243.512198763547
51.0106272140029 246.352628619061
51.2606793081892 245.203740338495
51.5107314023755 244.397316875789
51.7607834965618 245.252269083495
52.0108355907481 246.97961855043
52.2608876849344 248.084781994557
52.5109397791207 251.45315595404
52.7609918733069 255.822982892093
53.0110439674932 259.443938770089
53.2610960616795 262.563112786219
53.5111481558658 262.491424657938
53.7612002500521 265.23918807522
54.0112523442384 279.309879007542
54.2613044384247 287.58010188196
54.511356532611 302.693554363124
54.7614086267972 309.582277720123
55.0114607209835 314.11485095987
55.2615128151698 322.757278717634
55.5115649093561 329.652614884435
55.7616170035424 338.294408188795
56.0116690977287 343.25412474995
56.261721191915 346.527138101735
56.5117732861013 346.527138101735
56.7618253802876 346.527138101735
57.0118774744738 346.527138101735
57.2619295686601 346.527138101735
57.5119816628464 346.527138101735
57.7620337570327 346.527138101735
58.012085851219 346.527138101735
58.2621379454053 346.527138101735
58.5121900395916 346.527138101735
58.7622421337779 346.527138101735
59.0122942279642 346.527138101735
59.2623463221504 346.527138101735
59.5123984163367 346.527138101735
59.762450510523 346.527138101735
60.0125026047093 346.527138101735
60.2625546988956 346.527138101735
60.5126067930819 346.527138101735
60.7626588872682 346.527138101735
61.0127109814545 346.527138101735
61.2627630756408 346.527138101735
61.512815169827 346.527138101735
61.7628672640133 346.527138101735
62.0129193581996 346.527138101735
62.2629714523859 346.527138101735
62.5130235465722 346.527138101736
62.7630756407585 346.527138101735
63.0131277349448 346.527138101735
63.2631798291311 346.527138101735
63.5132319233174 346.527138101735
63.7632840175037 346.527138101735
64.0133361116899 346.527138101735
64.2633882058762 346.527138101735
64.5134403000625 346.527138101735
64.7634923942488 346.527138101735
65.0135444884351 346.527138101735
65.2635965826214 346.527138101735
65.5136486768077 346.527138101735
65.763700770994 346.527138101735
66.0137528651803 346.527138101735
66.2638049593665 346.527138101735
66.5138570535528 346.527138101735
66.7639091477391 346.527138101735
67.0139612419254 346.527138101735
67.2640133361117 346.527138101735
67.514065430298 346.527138101735
67.7641175244843 342.679168399289
68.0141696186706 341.455763411778
68.2642217128569 338.57729370685
68.5142738070431 336.137091294703
68.7643259012294 335.322202231154
69.0143779954157 332.633625092064
69.264430089602 331.322112337094
69.5144821837883 329.397166580092
69.7645342779746 328.699049095019
70.0145863721609 326.896719210493
70.2646384663472 324.921455760169
70.5146905605335 322.827118209547
70.7647426547197 319.76332846512
71.014794748906 334.17951455857
71.2648468430923 332.687886284241
71.5148989372786 329.49229449835
71.7649510314649 326.218644222876
};
\addlegendentry{actual consumption}
\end{axis}

\end{tikzpicture}

%% file: peak_red_eval.tex
% This file was created with tikzplotlib v0.10.1.
\begin{tikzpicture}

\definecolor{crimson2143940}{RGB}{214,39,40}
\definecolor{darkgray176}{RGB}{176,176,176}
\definecolor{darkorange25512714}{RGB}{255,127,14}
\definecolor{forestgreen4416044}{RGB}{44,160,44}
\definecolor{lightgray204}{RGB}{204,204,204}
\definecolor{steelblue31119180}{RGB}{31,119,180}

\begin{axis}[
width=250,
height=150,
legend cell align={left},
legend style={fill opacity=0.7, draw opacity=1, text opacity=1, draw=lightgray204},
tick align=outside,
tick pos=left,
x grid style={darkgray176},
xlabel={Percentage of Violations \(\displaystyle \Delta T_r\)},
xmin=-0.0894027777777778, xmax=1.98440277777778,
xtick style={color=black},
y grid style={darkgray176},
ylabel={Absolute PPR \(\displaystyle \Delta P_a\)},
ymin=28.5002884649398, ymax=47.3459537198711,
ytick style={color=black}
]
\small
\addplot [semithick, steelblue31119180, opacity=0.7, mark=*, mark size=2, mark options={solid}, only marks]
table {%
0.00486111111111111 40.7115659005369
};
\addlegendentry{alpha 0.001}
\addplot [semithick, darkorange25512714, opacity=0.7, mark=triangle*, mark size=2, mark options={solid,rotate=180}, only marks]
table {%
0.208055555555556 45.8137732070334
};
\addlegendentry{alpha 0.5}
\addplot [semithick, forestgreen4416044, opacity=0.7, mark=pentagon*, mark size=2, mark options={solid}, only marks]
table {%
0.690555555555556 46.4893325719197
};
\addlegendentry{alpha 1.0}
\addplot [semithick, crimson2143940, opacity=0.7, mark=square*, mark size=2, mark options={solid}, only marks]
table {%
1.89013888888889 29.3569096128912
};
\addlegendentry{baseline}
\end{axis}

\end{tikzpicture}

%% file: peak_red_var_k.tex
% This file was created with tikzplotlib v0.10.1.
\begin{tikzpicture}

\definecolor{darkgray176}{RGB}{176,176,176}
\definecolor{darkorange25512714}{RGB}{255,127,14}
\definecolor{forestgreen4416044}{RGB}{44,160,44}
\definecolor{lightgray204}{RGB}{204,204,204}
\definecolor{steelblue31119180}{RGB}{31,119,180}

\begin{axis}[
width=250,
height=150,
legend cell align={left},
legend style={
  fill opacity=0.8,
  draw opacity=1,
  text opacity=1,
  at={(0.03,0.97)},
  anchor=north west,
  draw=lightgray204
},
tick align=outside,
tick pos=left,
x grid style={darkgray176},
xlabel={Percentage of Violations \(\displaystyle \Delta T_r\)},
xmin=-0.0413784722222222, xmax=0.868947916666667,
xtick style={color=black},
y grid style={darkgray176},
ylabel={Absolute PPR \(\displaystyle \Delta P_a\)},
ymin=35.6080472918884, ymax=49.8745054177755,
ytick style={color=black}
]
\small
\addplot [semithick, steelblue31119180, opacity=0.7, mark=*, mark size=2, mark options={solid}, only marks]
table {%
0 36.2565226612469
};
\addlegendentry{12 timesteps}
\addplot [semithick, darkorange25512714, opacity=0.7, mark=triangle*, mark size=2, mark options={solid,rotate=180}, only marks]
table {%
0.690555555555556 46.4893325719197
};
\addlegendentry{36 timesteps}
\addplot [semithick, forestgreen4416044, opacity=0.7, mark=pentagon*, mark size=2, mark options={solid}, only marks]
table {%
0.827569444444444 49.226030048417
};
\addlegendentry{60 timesteps}
\end{axis}

\end{tikzpicture}

%% file: peak_red_var_bd.tex
% This file was created with tikzplotlib v0.10.1.
\begin{tikzpicture}

\definecolor{crimson2143940}{RGB}{214,39,40}
\definecolor{darkgray176}{RGB}{176,176,176}
\definecolor{darkorange25512714}{RGB}{255,127,14}
\definecolor{darkturquoise23190207}{RGB}{23,190,207}
\definecolor{forestgreen4416044}{RGB}{44,160,44}
\definecolor{lightgray204}{RGB}{204,204,204}
\definecolor{mediumpurple148103189}{RGB}{148,103,189}
\definecolor{steelblue31119180}{RGB}{31,119,180}

\begin{axis}[
width=250,
height=150,
legend cell align={left},
legend style={
  fill opacity=0.8,
  draw opacity=1,
  text opacity=1,
  at={(0.03,0.97)},
  anchor=north west,
  draw=lightgray204
},
tick align=outside,
tick pos=left,
x grid style={darkgray176},
xlabel={Percentage of Violations \(\displaystyle \Delta T_r\)},
xmin=0.534906684027778, xmax=0.714181857638889,
xtick style={color=black},
y grid style={darkgray176},
ylabel={Absolute PPR \(\displaystyle \Delta P_a\)},
ymin=51.2590887220473, ymax=63.7684850453046,
ytick style={color=black}
]
\small
\addplot [semithick, steelblue31119180, opacity=0.7, mark=*, mark size=2, mark options={solid}, only marks]
table {%
0.543055555555556 51.8276976458317
};
\addlegendentry{100 buildings}
\addplot [semithick, darkorange25512714, opacity=0.7, mark=triangle*, mark size=2, mark options={solid,rotate=180}, only marks]
table {%
0.61880787037037 55.3124733928164
};
\addlegendentry{120 buildings}
\addplot [semithick, forestgreen4416044, opacity=0.7, mark=pentagon*, mark size=2, mark options={solid}, only marks]
table {%
0.644990079365079 58.0627075933156
};
\addlegendentry{140 buildings}
\addplot [semithick, crimson2143940, opacity=0.7, mark=square*, mark size=2, mark options={solid}, only marks]
table {%
0.706032986111111 60.1604448936365
};
\addlegendentry{160 buildings}
\addplot [semithick, darkturquoise23190207, opacity=0.7, mark=diamond*, mark size=2, mark options={solid}, only marks]
table {%
0.666087962962963 61.8548542260543
};
\addlegendentry{180 buildings}
\addplot [semithick, mediumpurple148103189, opacity=0.7, mark=star, mark size=2, mark options={solid}, only marks]
table {%
0.638298611111111 63.1998761215202
};
\addlegendentry{200 buildings}
\end{axis}

\end{tikzpicture}

%% file: sol_time_res.tex
\begin{tikzpicture}

\definecolor{crimson2143940}{RGB}{214,39,40}
\definecolor{darkgray176}{RGB}{176,176,176}
\definecolor{forestgreen4416044}{RGB}{44,160,44}
\definecolor{lightgray204}{RGB}{204,204,204}
\definecolor{steelblue31119180}{RGB}{31,119,180}

\begin{axis}[
height=170,
legend cell align={left},
legend style={
  fill opacity=0.8,
  draw opacity=1,
  text opacity=1,
  at={(0.03,0.97)},
  anchor=north west,
  draw=lightgray204
},
log basis y={10},
tick align=outside,
tick pos=left,
width=452,
x grid style={darkgray176},
xlabel={Number of envelopes},
xmin=-0.6625, xmax=5.6625,
xtick style={color=black},
xtick={0,1,2,3,4,5},
xticklabels={50,100,250,500,1000,2000},
y grid style={darkgray176},
ylabel={Average solving time [s]},
ymajorgrids,
ymin=0.327609385578394, ymax=29.5454172746336,
ymode=log,
ytick style={color=black}
]
\small
\draw[draw=none,fill=crimson2143940,fill opacity=0.7] (axis cs:-0.375,0.000001) rectangle (axis cs:-0.125,0.426);
\addlegendimage{ybar,ybar legend,draw=none,fill=crimson2143940,fill opacity=0.7}
\addlegendentry{High request}

\draw[draw=none,fill=crimson2143940,fill opacity=0.7] (axis cs:0.625,0.000001) rectangle (axis cs:0.875,0.732);
\draw[draw=none,fill=crimson2143940,fill opacity=0.7] (axis cs:1.625,0.000001) rectangle (axis cs:1.875,1.826);
\draw[draw=none,fill=crimson2143940,fill opacity=0.7] (axis cs:2.625,0.000001) rectangle (axis cs:2.875,4.426);
\draw[draw=none,fill=crimson2143940,fill opacity=0.7] (axis cs:3.625,0.000001) rectangle (axis cs:3.875,9.873);
\draw[draw=none,fill=crimson2143940,fill opacity=0.7] (axis cs:4.625,0.000001) rectangle (axis cs:4.875,24.078);
\draw[draw=none,fill=steelblue31119180,fill opacity=0.7] (axis cs:-0.125,0.000001) rectangle (axis cs:0.125,0.402);
\addlegendimage{ybar,ybar legend,draw=none,fill=steelblue31119180,fill opacity=0.7}
\addlegendentry{Medium request}

\draw[draw=none,fill=steelblue31119180,fill opacity=0.7] (axis cs:0.875,0.000001) rectangle (axis cs:1.125,0.708);
\draw[draw=none,fill=steelblue31119180,fill opacity=0.7] (axis cs:1.875,0.000001) rectangle (axis cs:2.125,1.905);
\draw[draw=none,fill=steelblue31119180,fill opacity=0.7] (axis cs:2.875,0.000001) rectangle (axis cs:3.125,4.51);
\draw[draw=none,fill=steelblue31119180,fill opacity=0.7] (axis cs:3.875,0.000001) rectangle (axis cs:4.125,9.397);
\draw[draw=none,fill=steelblue31119180,fill opacity=0.7] (axis cs:4.875,0.000001) rectangle (axis cs:5.125,20.887);
\draw[draw=none,fill=forestgreen4416044,fill opacity=0.7] (axis cs:0.125,0.000001) rectangle (axis cs:0.375,0.408);
\addlegendimage{ybar,ybar legend,draw=none,fill=forestgreen4416044,fill opacity=0.7}
\addlegendentry{Low request}

\draw[draw=none,fill=forestgreen4416044,fill opacity=0.7] (axis cs:1.125,0.000001) rectangle (axis cs:1.375,0.688);
\draw[draw=none,fill=forestgreen4416044,fill opacity=0.7] (axis cs:2.125,0.000001) rectangle (axis cs:2.375,2.055);
\draw[draw=none,fill=forestgreen4416044,fill opacity=0.7] (axis cs:3.125,0.000001) rectangle (axis cs:3.375,4.037);
\draw[draw=none,fill=forestgreen4416044,fill opacity=0.7] (axis cs:4.125,0.000001) rectangle (axis cs:4.375,9.468);
\draw[draw=none,fill=forestgreen4416044,fill opacity=0.7] (axis cs:5.125,0.000001) rectangle (axis cs:5.375,21.819);
\end{axis}

\end{tikzpicture}